\documentclass[fleqn,usenatbib]{mnras}

\usepackage{newtxtext,newtxmath}
\usepackage[T1]{fontenc}
\usepackage{ae,aecompl}

\usepackage{graphicx}	% Including figure files
\usepackage{bm}
\let\vec\bm%

\usepackage{mleftright}
\let\left\mleft%
\let\right\mright%

\usepackage{amsmath}	% Advanced maths commands
\usepackage{amssymb}	% Extra maths symbols
\usepackage{etoolbox}
\makeatletter
\patchcmd\@combinedblfloats{\box\@outputbox}{\unvbox\@outputbox}{}{%
   \errmessage{\noexpand\@combinedblfloats could not be patched}%
}%
 \makeatother

\newcommand{\bs}[1]{\boldsymbol{#1}}
\newcommand{\be}{\begin{equation}}
\newcommand{\en}{\end{equation}}

\newcommand{\pder}[2]{\frac{\partial{#1}}{\partial{#2}}}

\newcommand{\der}[2]{\frac{d#1}{d#2}}

\newcommand{\Dx}{\Delta{}x}

\newcommand{\Dt}{\Delta{}t}
\newcommand{\f}{\frac}

\newcommand{\ex}{\vec{e}_x}
\newcommand{\ey}{\vec{e}_y}
\newcommand{\ez}{\vec{e}_z}

\newcommand{\un}{\vec{e}_n}
\newcommand{\um}{\vec{e}_m}
\newcommand{\up}{\vec{e}_p}

\newcommand{\para}{\parallel}

\newcommand{\paren}[1]{ \left( #1 \right) }
\newcommand{\kb}{k_{\textup {B}}}

\renewcommand{\b}{\bb{b}}

\newcommand{\dvz}{\delta \varv_{z}}

\newcommand{\mH}{m_{\textup{H}}}

\newcommand\bb[1]{\mbox{\boldmath{$#1$}}}
\newcommand\bcdot{\bb{\cdot}}
\newcommand\del{\bs{\nabla}}

\newcommand{\va}{\varv_\mathrm{a}}

\newcommand{\cs}{c}
\newcommand{\ca}{\tilde{c}}

\newcommand{\muG}{\umu\mathrm{G}}
\newcommand{\ui}{\mathrm{i}}
\newcommand{\ue}{\mathrm{e}}
\newcommand{\real}[1]{\mathrm{Re}(#1)}
\newcommand{\imag}[1]{\mathrm{Im}(#1)}

\title[Braginskii viscosity on a moving mesh]
{Braginskii viscosity on an unstructured, moving mesh \\
accelerated with super-time-stepping}

\author[Berlok, Pakmor and Pfrommer]{
Thomas Berlok$^{1}$\thanks{E-mail: tberlok@aip.de},
R\"udiger Pakmor$^{2}$ and
Christoph Pfrommer$^{1}$
\\
$^{1}$Leibniz-Institut f{\"u}r Astrophysik Potsdam (AIP),
 An der Sternwarte 16, D-14482 Potsdam, Germany \\
$^{2}$Max Planck Institute for Astrophysics, Karl-Schwarzschild-Stra{\ss}e 1,
 85748, Garching, Germany
}

\date{Accepted XXX. Received YYY; in original form ZZZ}

\pubyear{2019}

\usepackage{xcolor}

\begin{document}
\label{firstpage}
\pagerange{\pageref{firstpage}--\pageref{lastpage}}
\maketitle


\begin{abstract}
We present a method for efficiently modelling Braginskii viscosity on an
unstructured, moving mesh. Braginskii viscosity, i.e., anisotropic transport
of momentum with respect to the direction of the magnetic field, is thought to
be of prime importance for studies of the weakly collisional plasma that
comprises the intracluster medium (ICM) of galaxy clusters. Here anisotropic
transport of heat and momentum has been shown to have profound
consequences for the stability properties of the ICM. Our new method for
modelling Braginskii viscosity has been implemented in the moving mesh code
\textsc{Arepo}. We present a number of examples that serve to test the
implementation and illustrate the modified dynamics found when including
Braginskii viscosity in simulations. These include (but are not limited to)
damping of fast magneto-sonic waves, interruption of linearly polarized
Alfv\'{e}n waves by the firehose instability and the inhibition of the
Kelvin-Helmholtz instability by Braginskii viscosity. An explicit update of
Braginskii viscosity is associated with a severe time step constraint that
scales with $(\Delta x)^2$ where $\Delta x$ is the grid size. In our
implementation, this restrictive time step constraint is alleviated by
employing 2nd order accurate Runge-Kutta-Legendre super-time-stepping.
We envision including Braginskii viscosity in future
large-scale simulations of Kelvin-Helmholtz unstable cold fronts in cluster
mergers and AGN-generated bubbles in central cluster regions.
\end{abstract}


%-----------------------------------------------------------------------------
%-----------------------------------------------------------------------------
\section{Introduction} \label{sec:intro}

The present paper concerns itself with an extension of ideal
magnetohydrodynamics (MHD) which is known as Braginskii MHD \citep{Bra}. This
extension is appropriate for a fully ionized, weakly collisional and
magnetized plasma. The most prominent example of an astrophysical system with
these properties is the intracluster medium (ICM) of galaxy clusters.

The ICM has a high temperature ($T \sim 1-10$ keV) and low particle density
\citep[$n\sim 10^{-2}\,\mathrm{cm}^{-3}$, e.g.][]{Vik06}. This leads to a weak
collisionality as the Coulomb collision frequency has a $\sim n T^{-3/2}$
dependence on density and temperature \citep[e.g.][]{Hazeltine2004}. The ICM
also contains a magnetic field with a strength of $\sim 1\muG$ \citep{Car02}.
Although this magnetic field is dynamically weak, in the sense that the
thermal pressure is much larger than the magnetic pressure, it is strong
enough to magnetize the ICM, i.e., both ions and electrons have gyroradii that
are much smaller than the characteristic length scale of the plasma. In
summary, the temperature, density and magnetic field strength of the ICM lead
to the ordering \citep{schekochihin_turbulence_2006}
\be
     H \gg \lambda_{\mathrm{mfp}} \gg r_\mathrm{i} \gg r_\mathrm{e} \ ,
\en
where $H$ is a characteristic scale (e.g., the scale height),
$\lambda_{\mathrm{mfp}}$ is mean free path of ion Coulomb collisions,
$r_\mathrm{i}$ is the ion gyroradius and $r_\mathrm{e}$ is the electron
gyroradius.

As a consequence of this ordering, particles are tied to magnetic field lines
and travel long distances along them between each Coulomb collision. This
leads to heat and momentum transport which is primarily directed along the
local magnetic field direction. A fluid model for this anisotropic transport,
which differs from ideal MHD by just a few extra terms, can be constructed.
The extra terms describe anisotropic heat conduction and Braginskii viscosity
(also known as anisotropic viscosity) and the extension to ideal MHD including
both effects is known as Braginskii MHD \citep{Bra}. Simulations with
Braginskii MHD can show greatly modified dynamics compared to ideal MHD
simulations.

Anisotropic heat conduction fundamentally changes the stability properties of
ionized, weakly collisional, magnetized atmospheres (see \citealt{Bal16} for a
review).  The Schwarzschild criterion \citep{1958ses..book.....S}, which states
that an atmosphere is stable if the entropy increases with height, assumes a
collisional medium and has been found by \citet{Bal00,Bal01} and \citet{Qua08}
not to apply to weakly collisional atmospheres. They discovered that stability
instead depends on the direction of the temperature gradient and the orientation
of the magnetic field. Both temperature gradients anti-parallel and parallel to
gravity (increasing and decreasing with height) can be susceptible to
instability. The instabilities, known as the magneto-thermal instability (MTI,
\citealt{Bal00,Bal01}) and the heat-flux-driven buoyancy instability (HBI,
\citealt{Qua08}), are both driven by fast heat conduction along magnetic
fields. Depending on the magnetic field direction, anisotropic heat conduction
can therefore in principle render the outer parts of all clusters MTI-unstable
and the core regions of cool-core clusters HBI-unstable \citep{Vik06}.

The first studies of the MTI and the HBI considered anisotropic heat conduction
but neglected Braginskii viscosity because it acts on a slower time scale.
Nevertheless, the inclusion of Braginskii viscosity has subsequently been
shown to have important effects. \citet{Kun11} showed that the fastest growing
mode of the HBI has a vertical wavelength which is longer than the scale
height of the plane-parallel atmosphere. This made the local, linear
stability analyses utilized thus far inadequate and the theory was
consequently extended to a quasi-global setting in \citet{Lat12}. Similar
considerations and conclusions for the importance of including Braginskii
viscosity have been found for a generalization of the HBI in which both the
temperature and composition of the plasma is allowed to vary
\citep{Pes13,Berlok2015,Berlok2016b}.

One of the key questions which has been addressed by simulations of the HBI
is whether it will nonlinearly saturate to wrap the magnetic
fields around the cluster core, thus insulating the core from heat transport
from the hotter, outer regions \citep[see e.g.][]{Par08,Kun12}.
If such a re-orientation occurs, it is
important that it is accurately modeled by simulations. While simulations
without Braginskii viscosity find an efficient insulation, \citet{Kun12} found
that Braginskii viscosity prevents the HBI from reorienting the magnetic field
lines in all but the innermost regions. Besides opposing insulation of the
core by the HBI, Braginskii viscosity can also provide a heating mechanism for
the ICM via viscous heating \citep{kunz_thermally_2011}. Including Braginskii
viscosity in cluster simulations thus seems essential for understanding the
global energy budget of the ICM.

In general, simulations using Braginskii viscosity of the MTI/HBI, and their
composition generalizations mentioned above, show less small-scale mixing and
numerical reconnection \citep{Par12,Kun12,Berlok2016a,Berlok2016b}.
Simulations of the MTI also lead to a magnetic field structure with long
folds and sharp bends, a consequence of Braginskii viscosity suppressing
motions that change the magnetic field strength \citep{Kun12}.

Cavities that appear as depressions in plasma density have been observed in a
number of cool-core clusters \citep[e.g. Perseus, see ][]{Fabian2011}. The
cavities are believed to be AGN-inflated bubbles filled with hot, relativistic
plasma buoyantly rising in the ICM. Understanding whether such bubbles can
travel long distances into the ICM as well as how they eventually dissipate
their energy is an important ingredient in resolving the cooling-flow problem
\citep{Fabian1994}.  In disagreement with observations, however, bubbles in
idealized models are very quickly shredded by disrupting instabilities such as
the Rayleigh-Taylor instability and the Kelvin-Helmholtz instability (KHI), see
e.g. \citep{Churazov2001}.  Several stabilizing agents have been suggested in
order to make theory conform with observations, e.g., magnetic fields
\citep{Robinson2004,Ruszkowski2007}, \emph{isotropic} viscosity
\citep{Reynolds2005} or Braginskii viscosity \citep{Dong2009}.

\citet{Dong2009} showed that Braginskii viscosity, like magnetic fields,
suppress instabilities in the direction parallel to the magnetic field but
that motions in the perpendicular direction are uninhibited. The magnetic
field structure in the bubble region, which itself is likely regulated by
Braginskii viscosity, is therefore paramount in determining the stability of
bubbles \citep{Ehlert2018}.

Cold fronts are discontinuities in density and temperature found in galaxy
clusters (see \citealt{Markevitch2007,ZuHone2016} for reviews). In brief, cold
fronts are divided into two categories \citep{Tittley2005}: \emph{i)} remnant
core (or merger) cold fronts that arise during a merger when a smaller cluster
moves through the ICM of a larger cluster \citep{ZuHone2016} and \emph{ii)}
sloshing cold fronts which arise when low entropy gas from the central part of
a large cluster is displaced during a cluster merger \citep{Markevitch2007}.
Idealized computer models of sloshing cold fronts face an issue similar to the
one facing models of AGN-inflated bubbles: the KHI occurs with a vigorousness
unmatched by observations of real galaxy clusters. In sloshing cold fronts,
the two most promising stabilizing agents are Braginskii viscosity
\citep{ZuHone2015} and a magnetic field which has been shear-amplified to
sufficient strength at the location of the cold front \citep{ZuHone2011}.
Similarly, simulations have shown that Braginskii viscosity suppresses the KHI
in remnant core cold fronts \citep{Suzuki2013}. Whether the KHI is suppressed
by viscosity, magnetic field strength or some other agent is still an open
question \citep{ZuHone2016}.

The magneto-rotational instability (MRI, see \citealt{Balbus1998} for a review)
is a powerful mechanism for transporting angular momentum in accretion
discs. Initially discovered within the framework of ideal MHD
\citep{Balbus1991}, this instability exists also in weakly collisional
\citep{Balbus2004} and collisionless frameworks
\citep{Quataert2002,Sharma2003,Heinemann2014}.  Radiatively inefficient accretion
flows (RIAFs) are accretion discs found around super-massive black holes and
consist of extremely hot and dilute plasma (see \citealt{Quataert2003,Yuan2014}
for reviews). The characteristic temperatures and densities of RIAFs render them
even less collisional than the ICM and a proper treatment requires studying them
with methods applicable to collisionless plasmas.  Such studies, using
particle-in-cell (PIC) simulations, have however found that micro-scale
pitch-angle scattering of particle-waves induces an effective collision
frequency which is much higher than the Coulomb collision frequency
\citep{Kunz2014a,Hellinger2015b,Sironi2015,Kunz2016}. This has motivated studies
of RIAFs using Braginskii viscosity \citep[e.g.][]{Wu2017,Kempski2019} and even
relativistic variants thereof \citep{Chandra2017,Foucart2016,Foucart2017}, since
such models make it possible to study much larger time and length scales than
allowed by costly PIC simulations.

Besides the various applications to specific astrophysical systems, Braginskii
viscosity also, quite fundamentally, modifies the MHD wave family and their
propagation. It can be shown that the fast and slow magneto-sonic waves are
damped while circularly polarized Alfv\'{e}n waves are unmodified by
Braginskii viscosity (see e.g. \citealt{Par12} and our derivation of the
Braginskii MHD wave dispersion relation in Section~\ref{sec:wave-disp}). Only
recently, however, was it realized that linearly polarized Alfv\'{e}n waves
can be interrupted by the firehose instability in high-$\beta$ plasmas
\citep{Squire2016,Squire2017}. This finding has been shown to
have important consequences for the properties of turbulence in weakly
collisional plasmas, i.e., \citet{Squire2019} found in the limit of high
viscosity, that incompressible turbulent fluid motions are constrained to
move in a way in
which the magnetic field strength, $B$, does not change in time. Weakly
collisional plasmas with this interesting property are called
magneto-immutable in \citet{Squire2019}.

The rest of the paper is outlined as follows. In Section~\ref{sec:Equations}
we introduce
the equations of Braginskii MHD and derive the wave dispersion relation.
In Section~\ref{sec:Algo} we describe the
spatial discretization on an unstructured, moving mesh and the strategy for
updating the equations in time using Runge-Kutta-Legendre second order
accurate super-time-stepping. In Section~\ref{sec:tests} we provide a number
of tests including damping of fast magneto-sonic waves
(Section~\ref{sec:fast-waves}), the interruption of a linearly polarized
Alfv\'{e}n wave by the firehose instability (Section~\ref{sec:firehose}), and
the evolution of the KHI when a magnetic field is oriented along the flow
direction and Braginskii viscosity is taken into account
(Section~\ref{sec:KHI}). We conclude in Section~\ref{sec:conclusion} where we
also provide an outlook to the type of studies that we will be able to perform
in the future. Additional details about the \textsc{Arepo} implementation are
described in Appendices~\ref{sec:coord-trans} and \ref{app:local-equations},
while additional details about the analytical and numerical reference
solutions (used for testing the
implementation) are given in Appendices~\ref{sec:diff-solutions} and
\ref{sec:fft-calc}.

%-----------------------------------------------------------------------------
%-----------------------------------------------------------------------------
\section{Equations of Braginskii MHD}
\label{sec:Equations}
We introduce the equations of Braginskii magnetohydrodynamics (MHD) which
incorporate anisotropic
diffusion of heat and momentum
\citep{Bra,1983bpp..conf....1K,Sch10,Squire2017}.

In the following, $\rho$ is the mass density, $\vec{\varv}$ is the fluid
velocity, $\vec{B}$ is the magnetic field vector with magnitude $B$ and
direction $\b=\vec{B}/B$, $T$ is the temperature and $p$ is thermal pressure
given by $p = \rho \kb T/\mH \mu$ where $\kb$ is Boltzmann's constant, $\mH$
is the proton mass and $\mu$ is the mean molecular weight. The ratio of
thermal to magnetic pressure is given by $\beta=2\cs^2/\va^2$ where
$\cs=\sqrt{p/\rho}$ is the isothermal sound speed, $\va=B/\sqrt{\mu_0 \rho}$
is the Alfv\'{e}n speed and $\mu_0$ is the vacuum permeability. Finally, we
define the adiabatic sound speed as $\ca = \sqrt{\gamma p/\rho}$ where
$\gamma=5/3$ is the adiabatic index for a fully ionized, non-relativistic gas.

The mass continuity, momentum, induction and entropy equations are then given
by
\begin{align}
    \pder{\rho}{t} &= -\del \bcdot \left(\rho \vec{\varv}\right) \ ,
    \label{eq:rho}
\end{align}
\begin{align}
    \rho \der{\vec{\varv}}{t} &= - \del p -\del \bcdot \mathbf{\Pi}
    + \f{\left(\del \times \vec{B}\right)\times \vec{B}}{\mu_0}
    \ ,
    \label{eq:mom}
\end{align}
\begin{align}
    \pder{\vec{B}}{t} &= \del \times \left(\vec{\varv} \bs{\times}
    \vec{B}\right)
    \ ,
    \label{eq:ind}
\end{align}
and
\begin{align}
    \f{p}{\gamma -1}\der{\ln \left(p \rho^{-\gamma}\right)}{t} &=
    -\mathbf{\Pi} \vec{:} \del \vec{\varv} - \del \bcdot \vec{Q}\ ,
    \label{eq:ent}
\end{align}
where
\be
    \der{}{t} = \pder{}{t} + \left(\vec{\varv} \bcdot \del\right) \
    ,
\en
is the Lagrangian time derivative. In
Equations~\eqref{eq:rho}-\eqref{eq:ent}, the extra terms that are included in
Braginskii MHD (compared to the equations of ideal MHD) in order to model the
weak collisionality are the anisotropic heat flux, $\vec{Q}$, and the
anisotropic viscosity tensor, $\mathbf{\Pi}$.

The anisotropic heat flux is given by
\be
    \vec{Q} = - \chi_\para \b (\b \bcdot \del T) \ ,
\en
where $\chi_\para$ is the heat conductivity. This form of the heat flux has a
direction along the magnetic field and a magnitude which is proportional to
the gradient of temperature projected along the magnetic field. The heat
transport is
mainly carried by the electrons which can move long distances along the
magnetic field between collisions with other electrons. The heat conduction
will thus be primarily directed along the magnetic field. This form of
anisotropic heat conduction, which drives both the MTI and HBI, has already
been implemented in \textsc{Arepo} by \citet{Kannan2016}.

The anisotropic viscosity tensor is given by
\be
    \mathbf{\Pi} = - \Delta p \left(\b\b - \frac{\mathbf{1}}{3}\right) \ ,
    \label{eq:Pi}
\en
where the pressure anisotropy is defined as $\Delta p \equiv p_\perp -
p_\para$ and $\vec{b}\vec{b}$ is a dyadic product.
Here $p_\para$ ($p_\perp$) is the pressure parallel (perpendicular)
to the magnetic field and the total thermal pressure is $p = p_\para/3 + 2
p_\perp/3$.

Plasma motions can easily lead to the production of a pressure anisotropy due
to conservation of the first and second adiabatic invariants of plasma
particles \citep{Che56}. The resulting production of pressure anisotropy is
described by the Chew-Goldberger-Low (CGL) equations in a collisionless plasma
\citep{Che56,Baumjohann1996}. In a collisional plasma such a pressure
anisotropy is quickly isotropized by collisions. In a weakly collisional
plasma, where the collision frequency is much slower than the gyro-frequency
of particles but much faster than the dynamical frequency of the system, the
production of pressure anisotropy can be assumed to be balanced by collisions
\citep{schekochihin_plasma_2005}. This yields an expression for the pressure
anisotropy
\be
    \Delta p =
    \rho\nu_\para \der{\ln B^3 \rho^{-2}}{t}
     \label{eq:p-aniso}
     \ ,
\en
where $\nu_\para$ is the viscosity coefficient (a diffusion
coefficient with units m$^2/$s). We take the viscosity
coefficient to be Spitzer \citep{Spitzer1962,Bra,Sarazin1986} which means it
depends on density and temperature as\footnote{For Spitzer viscosity, the
viscosity coefficient is given by \citep{Hazeltine2004}
\be
    \nu_{\para} = \f{T}{\rho} \sum_\mathrm{s}
    \f{n_\mathrm{s}}{\nu_\mathrm{s}} \ ,
\en
where $n_\mathrm{s}$ is the number density and $\nu_\mathrm{s}$ is the
collision frequency of
ion species $\mathrm{s}$.} $\nu_\para \propto T^{5/2}/\rho$.
For constant density, we
can infer from Equation~\eqref{eq:p-aniso} that regions where the magnetic
field strength is increasing in time will have a positive pressure anisotropy
while
regions where the magnetic field strength is decreasing in time will have a
negative
pressure anisotropy. Both situations can lead to the excitation of microscale
instabilities if the magnitude of the pressure anisotropy exceeds certain
thresholds. A dominant perpendicular pressure can lead to the mirror
instability
\citep{Hasegawa1969,Southwood1993,Kivelson1996} while a dominant parallel
pressure can excite the firehose instability
\citep{Parker1958,Chandrasekhar1958,Rosenbluth1956,Vedenov1958,Davidson1968}.
These instabilities are not fully described by Braginskii MHD and require
special treatment (see Section~\ref{sec:firehose_and_mirror}).

Equation~\eqref{eq:p-aniso} can be rewritten by using the equation for mass
continuity, Equation~\eqref{eq:rho}, and the induction equation,
Equation~\eqref{eq:ind}, to rewrite the time derivatives of the mass density
and magnetic field strength. This yields\footnote{Here the notation $\vec{:}$
has the following meaning $\b\b \vec{:} \del \vec{\varv} = \sum_i \sum_j b_i
b_j \partial_i \varv_j$ which is equivalent to the trace of the matrix product
between the matrices $\b\b$ and $\del \vec{\varv}$.}
\be
    \Delta p =
    \rho\nu_\para\left(3\b\b\vec{:}\del \vec{\varv}
     -\del\bcdot\vec{\varv}\right) \ ,
     \label{eq:p-aniso-no-dt}
\en
showing that the pressure anisotropy can be evaluated from the instantaneous
velocity derivatives, and the density, temperature and local magnetic field
direction. Equation~\eqref{eq:p-aniso-no-dt} is used in \textsc{Arepo} when
evaluating the viscosity tensor.

The viscosity tensor enters in the momentum equation, Equation~\eqref{eq:mom},
and can be on the order of or larger than the Lorentz force when the magnetic
field is weak, i.e., $\beta>1$ \citep{Squire2017}. The viscosity tensor also
enters in the entropy equation as a viscous heating term
\be
    -\mathbf{\Pi} \vec{:} \del \vec{\varv} =
    \f{(\Delta p)^2}{3\rho \nu_\para} = \f{\rho \nu_\para}{3}
    \left(3\b\b\vec{:}\del \vec{\varv}
     -\del\bcdot\vec{\varv}\right)^2
    \ ,
    \label{eq:heating-term}
\en
which is always positive. The viscosity tensor, Equation~\eqref{eq:Pi}, has
vanishing bulk viscosity, i.e., the trace of the viscosity tensor is zero.

For our numerical implementation of Braginskii viscosity in \textsc{Arepo},
we will consider the evolution of the total energy density
(thermal + kinetic + magnetic)
\be
    E = \f{p}{\gamma - 1} + \f{1}{2}\rho \varv^2 + \f{B^2}{2\mu_0} \ ,
\en
instead of solving the entropy equation directly. The required evolution
equation for $E$ can be shown to be given by
\be
    \pder{E}{t} + \del \bcdot \left[ \paren{E+p_{\mathrm{T}}}\vec{\varv} -
    \f{\vec{B}\paren{\vec{B}\bcdot \vec{\varv}}}{\mu_0}\right]
    = -\del \bcdot \vec{Q}
      - \del \bcdot
    \paren{\vec{\Pi} \bcdot \vec{\varv}}
    \ , \\
    \label{eq:energy}
\en
where $p_{\mathrm{T}} = p + B^2/2\mu_0$ is the total pressure
(gas plus magnetic), and the advantage is that this approach conserves
total energy.

%-----------------------------------------------------------------------------
\subsection{Firehose and mirror instabilities}
\label{sec:firehose_and_mirror}

We consider two important microscale instabilities for the ions, namely the
firehose and the mirror instability. The firehose instability is excited if
\be
    \f{p_\perp}{p_\para} -1 < -\f{2}{\beta_\para} \ , \label{eq:firehose}
\en
and the mirror instability is excited if
\be
    \f{p_\perp}{p_\para} -1 > \f{1}{\beta_\para} \ , \label{eq:mirror}
\en
where $\beta_\para = 2\mu_0 p_\para/B^2$. These criteria for instability
can be rewritten into a criterion for stability \citep{Kun12}
\be
    \label{eq:fire_and_mirror}
    -\f{B^2}{\mu_0} < p_\perp - p_\para < \f{B^2}{2\mu_0} \ .
\en
Particle-in-cell simulations of the firehose and mirror instability indicate
that these instabilities saturate by driving the pressure anisotropy to
marginal stability \citep{Schekochihin2008,Rosin2011,Kunz2014a}, a prediction
that is also supported by observations of the solar wind
\citep{Bale2009,Chen2016}. Motivated by these studies we use an approach that
has also previously been used by e.g.
\cite{Sha06,Kun12,Berlok2016b,Squire2019}, i.e.,
we add an option to limit the value of the ion
pressure anisotropy to the limits set by the firehose and mirror instability
thresholds, Equation~\eqref{eq:fire_and_mirror}, when evaluating the viscosity
tensor.

%-----------------------------------------------------------------------------
%-----------------------------------------------------------------------------
\subsection{Dispersion relation for waves}
\label{sec:wave-disp}

We derive the dispersion relation for MHD waves which are modified when the
effect of Braginskii viscosity is taken into account.
In order to do so, we linearize
Equations~\eqref{eq:rho}-\eqref{eq:ent}. The only nonstandard term in the
linearization of these equations
is the viscosity term in the momentum equation\footnote{The viscosity term is
neglected in the linearized version of the internal energy equation because it
is second order in the perturbations}. We assume that $\Delta p = 0$ for the
equilibrium (as well as constant $\rho$ and zero background velocity). We then
find that
\be
    \delta \Delta p = i \rho \nu_\para
    \left(3\b\b\vec{:}\vec{k}\delta\vec{\varv} -
    \vec{k}\bcdot \delta \vec{\varv}\right) \ ,
\en
and the linearized equations are thus
\begin{align}
    -\ui\omega\f{\rho}{\rho} &= -\ui\vec{k} \bcdot  \delta \vec{\varv} \ , \\
    -\ui\omega \delta \vec{\varv} &= -\ui \vec{k} \f{\delta p}{\rho}
    -\vec{k}\bcdot\left[
    \nu_\para \left(3\b\b\vec{:}\vec{k}\delta\vec{\varv} -
    \vec{k}\bcdot \delta \vec{\varv}\right)
            \left(\b\b - \mathbf{1}/3\right)\right]
    \nonumber \\ &
    + \f{B^2}{\rho \mu_0}
    \left(\ui \vec{k} \times \delta \b\right)\times \b \ , \\
    -\ui \omega \delta \b &=
    \ui \vec{k} \times \left(\delta \vec{\varv} \times \b \right) \ , \\
    \f{\delta p}{p} &= \gamma \f{\delta \rho}{\rho} \ .
\end{align}
We then define the sound speed
\be
    \ca^2 \equiv \f{\gamma p}{\rho} \ ,
\en
and the Alfv\'{e}n speed
\be
    \va^2 \equiv \f{B^2}{\rho \mu_0} \ ,
\en
and combine the equations above to obtain a single equation for
$\delta \vec{\varv}$,
i.e.,
\begin{align}
    \label{eq:w^2-del-v}
    \omega^2 \delta \vec{\varv} &=
     \vec{k} \ca^2 \vec{k} \bcdot  \delta \vec{\varv}
    -\ui \omega \vec{k}\bcdot\left[
    \nu_\para \left(3\b\b\vec{:}\vec{k}\delta\vec{\varv} -
    \vec{k}\bcdot \delta \vec{\varv}\right)
    \left(\b\b - \mathbf{1}/3\right)\right]
    \nonumber \\ &
    + \va^2 \left[\vec{k} \times
    \left(\vec{k} \times
    \left(\delta \vec{\varv} \times \b \right)\right)\right]\times \b \ .
\end{align}
Assume now, without loss of generality, that $\b=\ez$ and $\vec{k}= k_\perp
\ex + k_\para \ez$. Equation~\eqref{eq:w^2-del-v} can then be written as
$\mathsf{D} \bcdot \delta \vec{\varv} = \vec{0}$ where $\mathsf{D} =
\mathsf{M}_{\mathrm{MHD}} + \mathsf{M}_{\mathrm{Brag}}$ and
\be
    \mathsf{M}_{\mathrm{MHD}}
    =
    \left(
    \begin{matrix}
    \omega^2 - k^2 \va^2 - k_\perp^2 \ca^2 & 0 & -\ca^2 k_\perp k_\para \\
    0 & \omega^2 - k_\para^2 \va^2 & 0 \\
    -\ca^2 k_\perp k_\para & 0 & \omega^2 - \ca^2 k_\para^2
    \end{matrix}
    \right) \ ,
\en
takes the standard MHD form (see e.g. \citealt{fitzpatrick2014plasma}) and
\be
    \mathsf{M}_{\mathrm{Brag}}
    =
    \ui \omega \f{\nu_\para}{3}
    \left(
    \begin{matrix}
    -k_\perp^2  & 0 & 2 k_\perp k_\para \\
    0 & 0 & 0 \\
    2 k_\perp k_\para & 0 & - 4 k_\para^2
    \end{matrix}
    \right) \ ,
\en
contains the Braginskii viscosity physics.

The determinant of $\mathsf{D}$ yields the dispersion relation for waves in
Braginskii MHD. We find, as for standard MHD, that the dispersion relation
factors into a dispersion relation for the shear Alfv\'{e}n wave
\be
    \left(\omega^2 - k_\para^2 \va^2\right) = 0 \ ,
    \label{eq:disp-shear-alfven}
\en
and one for the fast- and slow magneto-sonic waves
\begin{multline}
    \omega^4
    +\ui \omega^3 \f{\nu_\para}{3} \left(4k_\para^2 + k_\perp^2 \right)
    -\omega^2 k^2\left(\va^2 + \ca^2\right)
    \\ -\ui \omega \f{\nu_\para}{3} k_\para^2
        \left(9 k_\perp^2 \ca^2 + 4 k^2 \va^2\right)
    + k_\para^2 k^2 \ca^2 \va^2 = 0 \ .
    \label{eq:disp-sonic-waves}
\end{multline}
Equation~\eqref{eq:disp-shear-alfven} shows that shear Alfv\'{e}n waves
propagate without damping\footnote{The linearly polarized Alfv\'{e}n waves can
trigger the firehose instability and damp as a consequence
\citep{Squire2016,Squire2017} but this phenomenon is not captured by the
linear theory derived here.} in Braginskii MHD while
Equation~\eqref{eq:disp-sonic-waves} differs from the ideal MHD version by
terms that contain the viscosity coefficient, $\nu_\para$. Unlike the ideal
MHD result, Equation~\eqref{eq:disp-sonic-waves} contains odd powers of
$\omega$ which makes a simple solution for $\omega^2$ unattainable and the
analytical solution for $\omega$ difficult to analyze. We consider instead the
special limit $k=k_\perp$ used for studying damping of fast magnetosonic waves
in Section~\ref{sec:fast-waves}. We find in this limit that
Equation~\eqref{eq:disp-sonic-waves} becomes
\be
    \omega^2
    +\ui \omega \f{\nu_\para}{3} k_\perp^2
    - k_\perp^2\left(\va^2 + \ca^2\right)
    = 0 \ ,
    \label{eq:disp-sonic-perp}
\en
with solutions
\be
    \omega = \pm k_\perp \sqrt{\va^2 + \ca^2 -
            \left(\f{k_\perp\nu_\para}{6}\right)^2}
    - \ui \f{\nu_\para}{6} k_\perp^2 \ .
    \label{eq:fast-wave-damping}
\en
Braginskii viscosity thus adds an imaginary frequency which gives damping
at the rate $\nu_\para k_\perp^2/6$ as long as
\be
    k_\perp \nu_\para < 6 \sqrt{\va^2 + \ca^2} \ ,
    \label{eq:pure-damping-limit}
\en
and furthermore decreases the real frequency of the waves. When
Equation~\eqref{eq:pure-damping-limit} is violated at higher values of
$k_\perp \nu_\para$, the real part of the frequency disappears and $\omega$
becomes completely imaginary. The strong damping limit, i.e., when
$|\mathrm{Im}(\omega)|\geq |\mathrm{Re}(\omega)|$ occurs already when
\be
    k_\perp \nu_\para \geq 3\sqrt{2} \sqrt{\va^2 + \ca^2} \ .
    \label{eq:fast-damping-limit}
\en
Finally, the eigenmode structure of fast waves is found to be
$\delta b_z= \delta \rho/\rho=1$ and $\delta \varv_x = \omega_\pm/k_\perp$
where $\omega_\pm$ is given by Equation~\eqref{eq:fast-wave-damping}.
The results derived here have been used to test \textsc{Arepo} in
Section~\ref{sec:fast-waves}.

%-----------------------------------------------------------------------------
%-----------------------------------------------------------------------------
\section{Algorithm}
\label{sec:Algo}

We detail the implementation of Braginskii viscosity into the moving mesh code
\textsc{Arepo} which evolves fluid quantities, on an unstructured, moving
Voronoi mesh \citep{Springel2010,Pakmor2016a}. The implementation makes use of
\textsc{Arepo}'s MHD capabilities \citep{Pakmor2011,Pakmor2013} and is the
latest addition to a suite of non-ideal physics options in \textsc{Arepo}. These
options already include effects such as isotropic viscosity \citep{Munoz2013},
cosmic ray transport \citep{Pakmor2016,Pfrommer2017}, anisotropic heat
conduction \citep{Kannan2016}, Ohmic diffusion \citep{Marinacci2018} and
radiation transport \citep{Kannan2019}.

We have implemented Braginskii viscosity in \textsc{Arepo} by using operator
splitting to solve Equations~\eqref{eq:rho}-\eqref{eq:ind} and
Equation~\eqref{eq:energy}. The Braginskii viscosity step is added with a half
time step update before the main update of variables and a half time step
update after the main update.

Our algorithm thus solves only the viscous
terms in the energy and momentum equations given by
\begin{align}
    \label{eq:momentum-eq}
    \rho \pder{\vec{\varv}}{t} &= - \del \bcdot \mathbf{\Pi}, \\
    \label{eq:energy-eq}
    \pder{E}{t} &= - \del \bcdot \left(\mathbf{\Pi} \bcdot \vec{\varv}\right)
    \ ,
\end{align}
while the density, $\rho$, and the magnetic field is kept constant during the
Braginskii viscosity step. We use the equation for the evolution of the total
energy, Equation~\eqref{eq:energy}, instead of the equation for the entropy,
Equation~\eqref{eq:ent}, because this ensures that the algorithm conserves
energy.

%-----------------------------------------------------------------------------
\subsection{Spatial discretization on a Voronoi mesh}

\textsc{Arepo} uses an unstructured, moving Voronoi mesh for solving the
equations of motion \citep{Springel2010}. This makes the spatial
discretization of Equations~\eqref{eq:momentum-eq} and \eqref{eq:energy-eq}
non-trivial. We outline our strategy for evaluating the fluxes in the
following three subsections.

\subsubsection{A local coordinate system}
\label{sec:coordinate-systems}

We define the standard Cartesian coordinate unit vectors as $\ex$, $\ey$ and
$\ez$. We also define a coordinate system $\un$, $\um$ and $\up$ at each
interface between two Voronoi cells. This coordinate system is oriented such
that $\un$ is the normal to the interface between the two Voronoi cells.  The
unit vector $\um$ is parallel to the component of the magnetic field that lies
in the plane of the interface of the Voronoi cells. This is achieved by
defining $\um$ as\footnote{ This procedure breaks down if the
magnetic field is strictly normal to the interface, $\un=\b$. In this case the
directions of $\um$ and $\up$ are however also not important.}
\be
    \um = \f{\b - \left(\un\bcdot \b\right)
               \un}
               {|\b - \left(\un\bcdot \b\right)
               \un|} \ .
\en
Consequently, the magnetic field does not have a component along
$\up$ which is defined by
\be
    \up = \un \times \um \ .
\en
In this new coordinate system, the magnetic field is thus given simply by
$
\b = b_n \un + b_m \um
$
where $b_n \equiv \un \bcdot \b$ and $b_m = \um\bcdot \b$.
A general vector, such as the velocity, is written as
$ \vec{\varv} =
              \varv_n \un +
              \varv_m \um +
              \varv_p \up
$
in the local coordinate system. Finally, we also define the derivative
operators
    $\partial_n = \un \bcdot \del,$
    $\partial_m = \um \bcdot \del$
    and
    $\partial_p = \up \bcdot \del,$
where the gradient operator, $\del$, is given by
\be
    \del = \ex \pder{}{x} + \ey \pder{}{y} + \ez \pder{}{z} \ ,
\en
in the Cartesian coordinate system and by
\be
    \del = \un \partial_n + \um \partial_m + \up \partial_p \ ,
\en
in the local coordinate system. Coordinate transformations of vectors and
derivative operators are done in the usual way but are detailed in
Appendix~\ref{sec:coord-trans} for completeness.

\subsubsection{Finite volume equations in the local coordinate system}
\label{sec:spatial}
For finite volume codes such as \textsc{Athena} or \textsc{Arepo}
(which evolve volume averages of the physical
quantities inside grid cells) Equations~\eqref{eq:momentum-eq} and
\eqref{eq:energy-eq} are rewritten by taking the volume average
over a cell of volume $V$, i.e.,
\begin{align}
    \f{1}{V}\int_{V}\pder{(\rho \vec{\varv})}{t} \, dV &= - \f{1}{V}\int_{V}
    \del \bcdot \mathbf{\Pi}\, dV \ , \\
    \f{1}{V}\int_{V} \pder{E}{t} \, dV &= - \f{1}{V}\int_{V}
    \del \bcdot \left(\mathbf{\Pi} \bcdot \vec{\varv}\right) \, dV \ ,
\end{align}
which upon application of the divergence theorem become\footnote{ Here the
bars on the LHS of the equations denote volume averages. Such bars will not be
explicitly shown in the remainder of the paper, as they tend to clutter the
notation.}
\begin{align}
    \pder{\overline{\rho \vec{\varv}}}{t} &= - \f{1}{V}\int_{\partial V}
    \mathbf{\Pi} \bcdot \un \, dA \ ,
    \label{eq:momentum-flux} \\
    \pder{\overline{E}}{t} &= - \f{1}{V}\int_{\partial V}
    \left(\mathbf{\Pi} \bcdot \vec{\varv}\right) \bcdot \un \, dA \ .
    \label{eq:energy-flux}
\end{align}
where $\partial V$ denotes the surface of the volume and $dA$ is an
infinitesimal area. In our \textsc{Arepo} implementation, the surface
integrals on the RHS are approximated as a discrete sum of fluxes through the
faces of the Voronoi cell. The normal vectors, $\un$, are the ones described
in Section~\ref{sec:coordinate-systems}.

It is evident that we need to derive expressions for $\mathbf{\Pi} \bcdot \un$
and $\left(\mathbf{\Pi} \bcdot \vec{\varv}\right) \bcdot \un$. Using the
definition for the viscosity tensor, Equation~\eqref{eq:Pi}, we obtain
\be
\mathbf{\Pi} \bcdot \un = \Delta p \left(\b\b -\mathbf{1}/3\right)
    \bcdot \un =
    \Delta p \b b_n
    - \Delta p \un/3 \ .
    \label{eq:mom-flux-tmp}
\en
Since both $\b \bcdot \up = 0$ and $\un \bcdot \up=0$ (by construction),
Equation~\eqref{eq:mom-flux-tmp} shows that the viscous flux in the $\up$
direction is zero. This has the advantage that the $\varv_p$ component of the
velocity does not evolve when working in the local coordinate system.

We detail the conversion of Equations~\eqref{eq:momentum-eq} and
\eqref{eq:energy-eq} to the local coordinate system in
Appendix~\ref{app:local-equations} where we find that the momentum and energy
equations can be written as\footnote{The mass density stays constant during a
Braginskii viscosity step and can therefore be pulled out from the time
derivative in the momentum equations.}
\begin{align}
    \rho \pder{\varv_n}{t} &= - \f{1}{V}\int_{\partial V}
    \Delta p \left(b_n b_n - 1/3\right) \, dA \ ,
    \label{eq:v_n}
    \\
    \rho \pder{\varv_m}{t} &= - \f{1}{V}\int_{\partial V}
    \Delta p\, b_m b_n \, dA \ ,
    \label{eq:v_m}
    \\
    \pder{E}{t} &= - \f{1}{V}\int_{\partial V}
    \Delta p \left(b_n b_n \varv_n + b_n b_m \varv_m - \varv_n/3 \right)
            \, dA \ ,
    \label{eq:E}
\end{align}
where the pressure anisotropy is given by
\begin{align}
    \Delta p =
    3\rho \nu_\para \left[
    b_n^2 \partial_n \varv_n +
    b_m^2 \partial_m \varv_m +
    b_n b_m \left(\partial_m \varv_n + \partial_n \varv_m \right)
    \right] \nonumber \\
    -\rho \nu_\para \left(\partial_n \varv_n + \partial_m \varv_m + \partial_p
    \varv_p\right) \ .
    \label{eq:p_aniso_loc}
\end{align}

\subsubsection{Estimating derivatives and values at cell interfaces}

All quantities and their derivatives are estimated at the corners of
Voronoi cells by employing the procedure outlined in Section 2.1 of
\citet{Pakmor2016}, i.e. by performing a least-squares fit using values at
the three (four) adjacent cell centers in two (three) dimensions.

A scalar quantity, $\phi$, is then estimated at the interface of a Voronoi
cell by performing a
weighted mean
\be
    \phi_{\mathrm{face}} = \sum_i w_i \phi_i \ ,
\en
where $w_i$ is the weight of the corner $i$.\footnote{As described in
\citet{Pakmor2016}, the least-squares fit can fail if the mesh contains
highly irregular cells. Failed corners are assigned a zero weight and the
weights of the remaining corners are adjusted accordingly. We do not calculate
the viscous flux across an interface when it has less than two good
corners. Mesh regularization in \textsc{Arepo} ensures that such failed
interfaces rarely occur. We test the implementation on an irregular, random
grid in Section~\ref{sec:fast-waves}.}

A derivative, for instance $\partial\phi/\partial x$, could in principle also
be estimated by using a similar weighting procedure, e.g.,
\be
    \pder{\phi_{\mathrm{face}}}{x} = \sum_i w_i \pder{\phi_i}{x} \ .
\en
Experience has however shown that more care is needed for anisotropic
transport. This was originally realized by \citet{Sha07} for anisotropic heat
conduction where it was shown that heat sometimes flows from cold to hot when
a simple averaging procedure is used. For Braginskii viscosity, a similar
issue is that a simple average can lead to flow of momentum in the wrong
direction \citep{Par12,ZuHone2015}. As this issue arises when not all
estimates agree on the sign of the derivative, our procedure for estimating
the derivative at interfaces is the following: if all the corners of an
interface agree on the sign of the derivative, we estimate the derivative at
the interface by calculating a weighted harmonic mean of the values at corners

\be
    \pder{\phi_{\mathrm{face}}}{x} = \left(\sum_i \f{w_i}{\partial
    \phi_i/\partial x}\right)^{-1} \ .
    \label{eq:limited-derivative}
\en
If the signs disagree, we set the estimate to zero. In two dimensions (where
$w_i =1/2$), Equation~\eqref{eq:limited-derivative} is simply the Van Leer
limiter which is also used in the Braginskii viscosity implementation in
\textsc{Athena} \citep{stone_athena:_2008,Par12}.

%-----------------------------------------------------------------------------
\subsection{Second order accurate super-time-stepping}
\label{sec:STS}

Stability consideration imply a time step constraint for an explicit update of
viscosity,
\be
    \Dt \leq C \f{(\Delta x)^2}{2 d \nu_\para} \ ,
    \label{eq:dt-brag}
\en
where $\Delta x$ is minimum size of the Voronoi cells, $d$ is the dimension of
the simulation and $C$ is the Courant number. In comparison, the MHD time step
constraint is
\be
    \Dt_\mathrm{MHD} \leq C \f{\Dx}{\varv_{\mathrm{max}}} \ ,
    \label{eq:dt-MHD}
\en
where $\varv_{\mathrm{max}}$ is the maximum signal speed (flow velocity plus
the velocity of the fast magnetosonic wave). Due to the different scalings
with $\Dx$, the MHD time step constraint will thus often be such that
$\Dt_\mathrm{MHD}\gg\Dt$, in particular for high resolution simulations.

In order to alleviate this constraint, we have implemented a second order
accurate super-time-stepping (STS) version of Braginskii viscosity. This
implementation is based on \citet{Meyer2012,Meyer2014} which describe a
second-order accurate Runge-Kutta Legendre method (RKL2). The RKL2 method has
been implemented in the Pluto code \citep{Vaidya2017} for anisotropic heat
conduction but has not previously been used for anisotropic viscosity. We
briefly outline the merits of STS and how we have applied it to Braginskii
viscosity.

Super-time-stepping introduces several new stages for each super time step.
The main advantage is that the maximum time step, $\tau$, is given by
\be
    \tau = \f{\Dt}{4}\left(s^2 + s - 2\right) \ ,
    \label{eq:sts-tau}
\en
where $\Dt$ is the maximum time step for an explicit update and $s$ is the
number of stages in the super-time-stepping. For large $s$, $\tau\propto s^2$
and a speedup by a factor $s$ is therefore expected compared to subcycling of
an explicit update.

In Braginskii viscosity, the momentum equation is a diffusion equation which
has a diffusion coefficient that depends on temperature. The viscous heating
in the energy equation can change the temperature and thus the viscosity
coefficient. The equations are thus coupled. In equation form, we can
write
\begin{align}
    \pder{\vec{\varv}}{t} &= \vec{L}\left(T, \vec{\varv}\right) \ , \\
    \pder{E}{t} &= \del \bcdot \vec{F}_E(T, \vec{\varv}) \ ,
\end{align}
where $\vec{L}(T, \vec{\varv})$ is the parabolic operator on the RHS of
Equation~\eqref{eq:momentum-eq} and $\vec{F}_E(T, \vec{\varv})$ is the energy
flux on the RHS of Equation~\eqref{eq:energy-eq}. Both equations depend on
both velocity and temperature. In order to solve the momentum equation using
super-time-stepping, we decouple the momentum and energy equations by assuming
that the viscosity coefficient simply takes the value at the beginning of the
time step. This allows us to write the velocity update from time $t^n$ to time
$t^{n+1}=t^n + \tau$ as \citep{Meyer2012}
\begin{align}
    \vec{Y}_0 =&\; \vec{\varv}^n \ , \\
    \vec{Y}_1 =&\; \vec{Y}_0 + \tilde{\mu}_1 \tau \vec{L}(T^n, \vec{Y}_0) \ , \\
    \vec{Y}_j =&\; \mu_j \vec{Y}_{j-1} + \nu_j \vec{Y}_{j-2} + (1 - \mu_j - \nu_j) \vec{Y}_0 \nonumber
    \\
    & + \tilde{\mu}_j \tau \vec{L}(T^n, \vec{Y}_{j-1}) +
    \tilde{\gamma}_j \tau \vec{L}(T^n, \vec{Y}_0)  \quad \mathrm{for} \: 2\leq j \leq s
    \\
    \vec{\varv}^{n+1} =&\; \vec{Y}_s \ .
\end{align}
Here the coefficients $\tilde{\mu}_1$, $\mu_j$, $\nu_j$, $\tilde{\mu}_j$
$\tilde{\gamma}_j$ are the super-time-stepping parameters\footnote{We retain
their notation but stress that these parameters should not be confused with
physical constants such as viscosity, adiabatic index etc.} given in equations
16-17 in \citet{Meyer2012}.

In order to also update the total energy, we use the trapezoidal
rule, i.e.,\footnote{An alternative
to this approach would be to update the
total energy by performing an Euler step for each stage of the
super-time-step, i.e.,
\be
E_{j+1} = E_{j} + \delta t_j
 \del \bcdot \vec{F}_E\left(T^j, \vec{\varv}^{j}\right)
 \quad \mathrm{for} \: 0\leq j < s \ ,
\en
where the time increment for each stage is given by $\delta t_0 = \Dt/3$,
$\delta t_1 = 2\Dt/3$, $\delta t_j = (j+1) \Dt/2$ for $j>1$
and fullfill $\sum_j \delta t_j = \tau$.
This would allow for the viscosity coefficient to
vary during the super-time-step and could potentially provide an increased
accuracy when using many stages.}
\be
    E^{n+1} = E^n + \f{\tau}{2}
    \left[\del \bcdot \vec{F}_E\left(T^n, \vec{\varv}^{n}\right)
      + \del \bcdot \vec{F}_E\left(T^n, \vec{\varv}^{n+1}\right)
    \right] \ .
\en
The number of super-time-stepping stages needed for each time step is likely
to change as a function of time. In the current implementation we determine
the number of necessary stages by first computing the minimum values of $\Dt$
and $\Dt_{\mathrm{MHD}}$ on the grid. We then compute $s$ such that $\Dt(s^2 +
s -2)\geq \Dt_{\mathrm{MHD}}$ with $3 \leq s \leq s_{\mathrm{max}}$ where
$s_{\mathrm{max}}$ is a pre-defined maximum number of stages. We only use odd
values for $s$ as recommended by \citet{Meyer2012}.

%-----------------------------------------------------------------------------
%-----------------------------------------------------------------------------
\section{Tests and examples}
\label{sec:tests}

%-----------------------------------------------------------------------------
\subsection{Decay of a velocity profile I}
\label{sec:decay1}

\begin{figure*}
    \includegraphics[trim= 0 25 0 0]{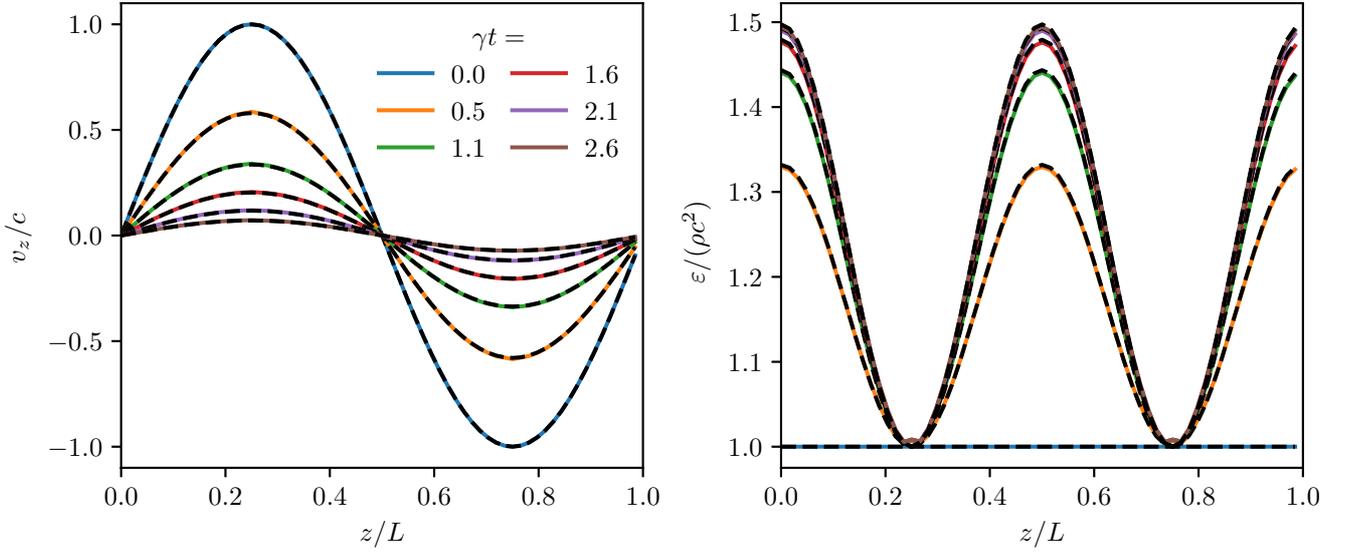}
    \caption{
    Viscous damping of a velocity profile (left panel) and the associated
    increase in internal energy due to viscous heating (right panel). The data
    from the 3D simulation (solid lines) agrees with the analytical solution
    (black dashed lines).
    }
    \label{fig:3D-Brag}
\end{figure*}

We test the numerical method for solving Equations~\eqref{eq:momentum-eq} and
\eqref{eq:energy-eq} by using a simple initial condition for which we can also
derive the analytical solution. The MHD solver is turned off for this test and
the magnetic field is static.
We consider an initial condition with velocity
\be
    \vec{\varv}(\vec{r}) = A \sin(\vec{k} \bcdot \vec{r})\f{\vec{k}}{k} \ ,
\en
a uniform magnetic field, $\vec{B}$, and an initially constant internal
energy, $\varepsilon_0$. Here $A$ is the amplitude of the perturbation,
$\vec{r}=x\ex+y\ey+z\ez$ is a position vector and $ \vec{k}$ is a wavevector
with magnitude $k$. If the magnetic field, $\vec{B}$, wave vector, $\vec{k}$,
and velocity, $\vec{\varv}$, are aligned, the time evolution of velocity and
internal energy can be shown to be given by\footnote{A derivation of this
solution is presented in Appendix~\ref{sec:1d-diff-solution}.}
\begin{align}
    \vec{\varv}(\vec{r}, t) &= A \sin(\vec{k} \bcdot \vec{r})e^{-\gamma t}
    \f{\vec{k}}{k} \ , \\
    \varepsilon(\vec{r}, t) &= \varepsilon_0 + \f{\rho A^2}{2} \cos^2(\vec{k}
    \bcdot
    \vec{r}) \left(1 - \ue^{-2\gamma t}\right) \ ,
\end{align}
where
\be
    \gamma = \f{4\nu_\para}{3}k^2 \ ,
\en
is the decay rate.

We perform a simulation in a three-dimensional (3D) periodic, cubic box of
size $L$ on a static Voronoi mesh with $2N^3$ mesh-generating points
where $N=32$. This mesh is created by interlacing two Cartesian meshes, i.e.,
with the first mesh displaced to have its cell centers at the cell corners of
the second mesh.
The magnetic field, wave vector and velocity field are all
oriented along $\ez$ but nothing is aligned with the Voronoi grid.
This means that the simulation
is automatically testing all the terms in the implementation. We use STS with
$s=31$ stages to accelerate the simulation, yielding a speedup by a factor
$\sim 8$. The result is shown in Fig.~\ref{fig:3D-Brag} and shows excellent
agreement with the analytical solution. Physically, the viscous heating is
maximal at the nodes of the velocity profile because the slope of the velocity
profile is maximal there.

%-----------------------------------------------------------------------------
\subsection{Decay of a velocity profile II}
\label{sec:decay2}

\begin{figure*}
    \includegraphics[trim= 0 25 0 10]{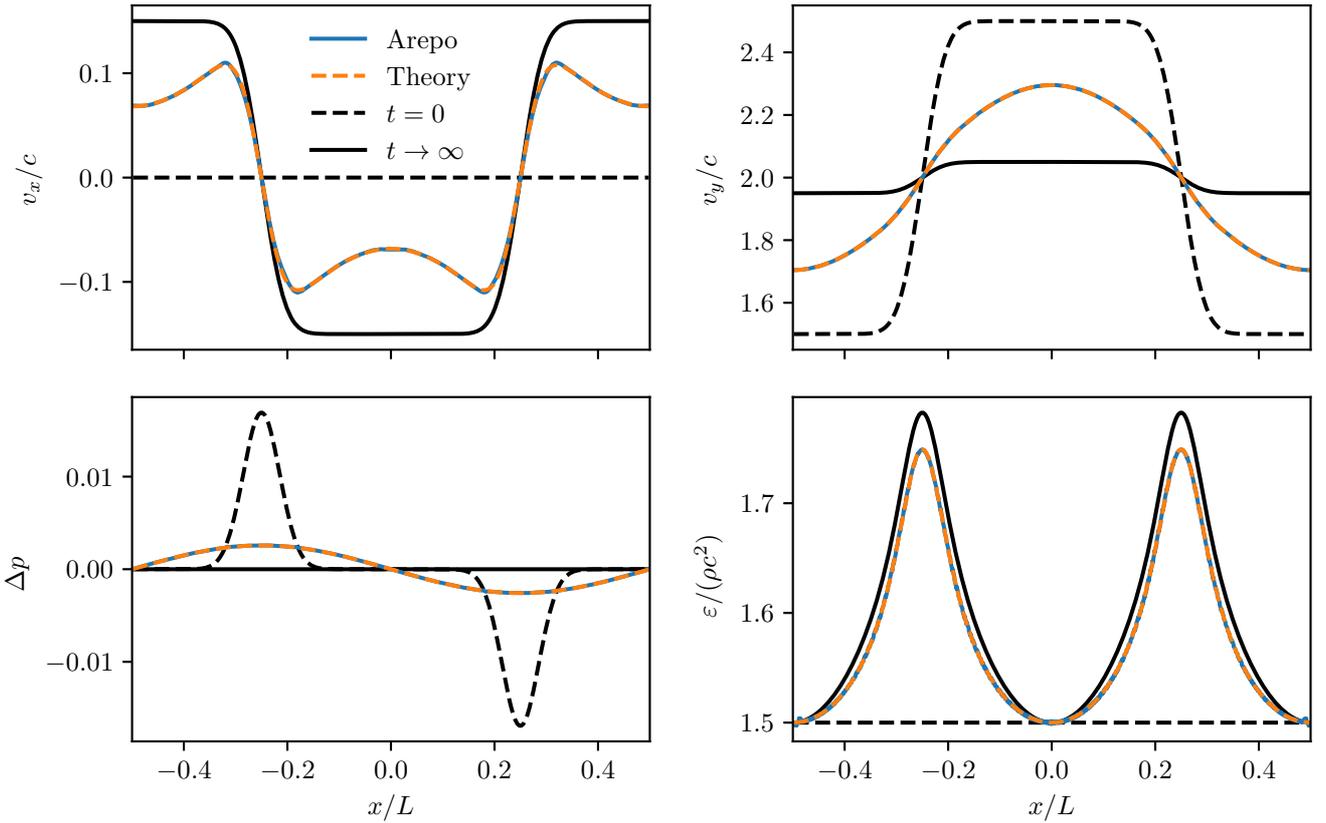}
    % trim={<left> <lower> <right> <upper>}
    \caption{
    Anisotropic diffusion in a magnetic field $\vec{B} = B(\ex + \ey)/
    \sqrt{2}$ creates a non-zero $\varv_x$ (upper left panel) from a velocity
    field which is initially strictly in the $y$-direction (upper right
    panel). The pressure anisotropy (lower left panel) causes viscous heating
    which modifies the internal energy (lower right panel). We compare data
    from \textsc{Arepo} (blue solid lines) and analytical theory (orange
    dashed lines) at $\cs t/L = 25$. The initial condition ($\cs t/L = 0$) and
    the asymptotic solution ($\gamma_n t \gg 1$) are shown with black dashed
    and black solid lines, respectively.
    }
    \label{fig:hopkins}
\end{figure*}

We consider a simulation inspired by the test problem in \citep{Hopkins2017}.
We take $\vec{B} = B(\ex + \ey)/\sqrt{2}$ with an initial velocity profile
given by $\vec{\varv} = \cs q(x) \ey$ where
\be
    q(x) = \f{3}{2} -\f{1}{2} \left(\mathrm{erf}\left(\f{x-x_0}{a}\right)
                                    - \mathrm{erf}\left(\f{x+x_0}{a}\right)
                                    \right) \ ,
\en
with $x_0 = 1/4L$ and $a=0.05L$. Unlike our first example in
Section~\ref{sec:decay1}, Equations~\eqref{eq:momentum-eq} and
\eqref{eq:energy-eq} cannot be reduced to a simple one-dimensional diffusion
equation for this setup and we instead have a coupled set of equations. This
set of equations can however still be solved analytically and we present a
derivation of the analytical solution in Appendix~\ref{sec:2d-diff-solution}.
Our derivation uses the fact that $q(x)$ can be represented by the Fourier
series
\be
    q(x) = \sum_{n=0}^{n=\infty} a_n \cos(k_n x) \ ,
\en
where $k_n = 2\pi n /L$ and the Fourier coefficients are given by
\be
    \label{eq:fourier-an}
    a_n =\left\{ \begin{array}{cc}
    2 &\mathrm{for}\; n=0 \ , \\
    -2\f{\sin(3 n\pi/2)}{n\pi} \ue^{-n^2 \pi^2/400} &\mathrm{for}\; n>0 \ .
    \end{array}\right.
\en
We find that the evolution of $\varv_x$ and $\varv_y$ can be described by
\begin{align}
    \varv_x(x, t) &= -\cs\sum_{n=0}^{\infty} \f{3a_n}{10} \cos(k_n x)
    \left(1 - \ue^{-\gamma_n t} \right) \ ,
    \label{eq:vx-h-sol}
\end{align}
and
\begin{align}
    \varv_y(x, t) &= \cs\sum_{n=0}^{\infty} \f{a_n}{10} \cos(k_n x)
    \left(1 + 9 \ue^{-\gamma_n t}\right)
    \label{eq:vy-h-sol} \ ,
\end{align}
where
\be
    \gamma_n = \f{5\nu_\para}{6}k_n^2 \ .
\en
We conclude from the time dependence of Equations~\eqref{eq:vx-h-sol} and
\eqref{eq:vy-h-sol} that $\varv_x=0$ at $t=0$ but that a non-zero component
will grow as a function of time. This happens at the expense of $\varv_y$
which decreases in time. The mixing of $\varv_x$ and $\varv_y$ happens because
the magnetic field is misaligned with respect to the initial velocity profile.
Interestingly, the asymptotic solution with $\gamma_n t \gg 1$ has both
$\varv_x$ and $\varv_y$ vary as a function of the $x$-coordinate. For
isotropic viscosity both velocity components would be constant in space after
a long duration of time. For Braginskii viscosity, things are not so simple.
Instead, the solution at $\gamma_n t \gg 1$ is such that a particular linear
combination of $\varv_x$ and $\varv_y$ is constant in space, i.e., one can see
from Equations~\eqref{eq:vx-h-sol} and \eqref{eq:vy-h-sol} that $\varv_x + 3
\varv_y = 6 \cs$ at late times. This follows from the geometry of the problem
and is outlined in more detail in Appendix~\ref{sec:diff-solutions}. We
conclude that velocity profiles can change in complicated ways in the presence
of anisotropic Braginskii viscosity.

The velocity profiles yield a pressure anisotropy given by
\be
    \Delta p(x, t) = -\f{3\rho \cs \nu_\para}{2} \sum_{n=1}^{\infty} k_n a_n
    \sin(k_n x) \,
    \ue^{-\gamma_n t} \ ,
    \label{eq:pa-h-sol}
\en
and the effect of Braginskii viscosity is to drive this pressure anisotropy
towards zero, as seen by considering the $\gamma_n t \gg 1$ limit of
Equation~\eqref{eq:pa-h-sol}.

The viscous damping has an associated heating
rate that will increase the internal energy in regions where $\Delta p \neq
0$.  We obtain an expression for the
evolution of the internal energy given by
\begin{multline}
   \varepsilon(t) = \varepsilon_0 + \f{9\rho \cs^2 }{10}
   \sum_{n=1}^{\infty}\sum_{m=1}^{\infty} a_n a_m \f{\sqrt{\gamma_n
   \gamma_m}}{\gamma_n+\gamma_m}
   \times  \\
   \sin(k_n x) \sin(k_m x)
   \left(1 - e^{-(\gamma_n+\gamma_m) t}\right) \ ,
   \label{eq:eps-h-sol}
\end{multline}
where $\varepsilon_0$ is the initially constant internal energy. As
we might
have been able to conclude already from the evolution of the pressure
anisotropy, we observe that the internal energy stops increasing and becomes
constant in time as $\gamma_n t \gg 1$.

We can use the analytical solution given by Equations~\eqref{eq:vx-h-sol},
\eqref{eq:vy-h-sol}, \eqref{eq:pa-h-sol}, and \eqref{eq:eps-h-sol} to test our
implementation of Braginskii viscosity.  In order to do so, we construct a 2D
grid with resolution $N_x=128$ and $N_y=4$ and size $L_x = L = L_y N_x/N_y$.
The Voronoi cell centers are set to be a regular Cartesian, uniform grid where
every second row has been displaced by $0.45 \Dx$, thus creating a hexagonal
mesh \citep{Pakmor2016}. We set $\nu_\para/(L \cs)=10^{-3}$ and evolve the
simulation to $\cs t/ L= 25$ with the MHD solver turned off.\footnote{In
\citet{Hopkins2017} a similar test was performed with the MHD solver turned
on, with the justification that MHD dynamics would be sub-dominant with a
magnetic field strength of $B_0=10^{-6}$ in code units. While this argument is
valid for ideal MHD, Braginskii MHD is susceptible to microscale instabilities
when Equation~\eqref{eq:fire_and_mirror} is violated. These appear to have
modified the solution presented in \citet{Hopkins2017} which exhibits an
asymmetry around $x=0$. As apparent in Fig.~\ref{fig:hopkins}, the initial
condition has a significant pressure anisotropy such that the mirror
(firehose) instability threshold is violated at $x/L = -1/4$ ($x/L =1/4$). We
therefore turn off the MHD solver in \textsc{Arepo} which prevents microscale
instabilities from modifying the evolution of the simulation. We have also
performed the test with \textsc{Athena} (not shown here) with the MHD solver
turned off and find agreement between \textsc{Athena}, \textsc{Arepo} and the
analytical solution. } The resulting profiles for $\varv_x$, $\varv_y$,
$\Delta p$ and $\varepsilon$ are shown in Fig.~\ref{fig:hopkins} with
\textsc{Arepo} data as blue solid lines and the theory as orange dashed lines.
The initial condition and the $\gamma_n t \gg 1$ analytical solution are also
shown with black dashed lines and black solid lines, respectively. As expected
from the analytical solution, we observe that the $\varv_x$-component grows
(upper left panel) while the initial $\varv_y$-component decays (upper right
panel). The pressure anisotropy (lower left panel), which initially has two
sharp peaks of opposite sign decreases as a function of time. At $\cs t/ L=
25$ most of the high wavenumber modes have already decayed and the pressure
anisotropy looks almost like the fundamental sine wave (which decays at the
slowest rate). The internal energy has developed two prominent peaks at the
locations where the pressure anisotropy was initially peaked ($x/L=\pm 1/4$)
and the internal energy has almost achieved its final value.

%-----------------------------------------------------------------------------
\subsection{Fast magnetosonic modes}
\label{sec:fast-waves}

\begin{figure*}
    \includegraphics{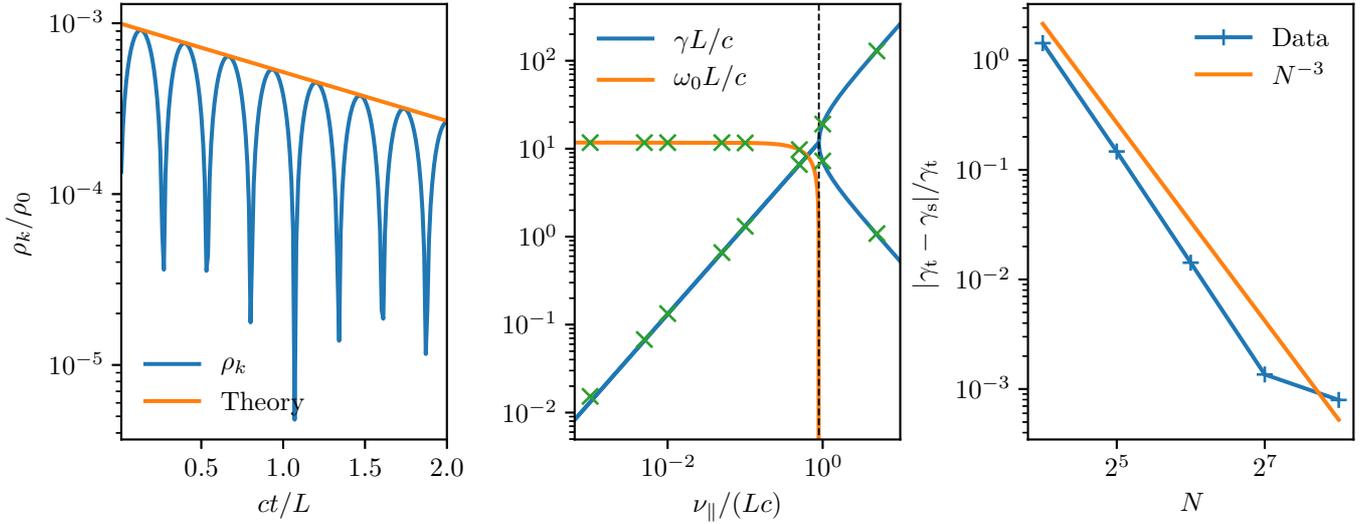}
    \caption{
    Damping of fast magnetosonic waves by Braginskii viscosity. Left:
    the density amplitude both oscillates and decays when $\nu_\para/(L\cs) =
    0.05$. Middle: the damping rate, $\gamma$, and real frequency,
    $\omega_0$, as measured in simulations with varying viscosity coefficient,
    $\nu_\para$. We compare with the theoretical prediction given by
    Equation~\eqref{eq:fastwave-damping-main}. The discrepancy at low values
    of $\nu_\para$ is due to numerical viscosity. Modes with $\nu_\para$ to
    the right of the vertical dashed line have $\omega_0=0$ and suffer pure
    decay. Right: fractional error in the damping rate as a function of
    grid resolution, $N=N_x=N_y$. As in \citet{Par12}, we obtain third order
    convergence.
    }
    \label{fig:fast-waves}
\end{figure*}

We consider a 2D domain in the $xy$-plane of size $[0,L]\times[0, L]$ with a
background magnetic field perpendicular to this plane, i.e. $\vec{B} = B_0
\ez$ where $B_0$ is such that $\beta=25$.
We initialize a standing, fast magnetosonic mode with initial
velocity, density and magnetic field given by the linear wave solution
which we derived in Section~\ref{sec:wave-disp}, i.e.,
\begin{align}
    \vec{\varv}(\vec{r}, t) &= -A \sin(\vec{k}\bcdot \vec{r})
    \left[\omega_0 \cos(\omega_0 t) - \gamma
    \sin(\omega_0 t)\right] \ue^{-\gamma t}
    \f{\vec{k}}{k^2}
    \label{eq:fastwave-eig-v} \ , \\
    \f{\delta \rho}{\rho_0} &= \f{\delta B_z}{B_0} =
    A \cos(\vec{k}\bcdot \vec{r})\sin(\omega_0 t)\,\ue^{-\gamma t} \ ,
    \label{eq:fastwave-eig-B}
\end{align}
where $\vec{k} = k_x \ex + k_y \ey$ with $k_x=k_y=k_\perp/\sqrt{2} = 2\pi/L$
and $A=10^{-3}$ is the perturbation amplitude. Here $\omega_0=\real{\omega}$
and $\gamma=-\imag{\omega}$ are the real and imaginary parts of the complex
frequency, $\omega$, which is given by
\be
    \omega = \pm k_\perp
    \sqrt{\va^2 + \ca^2 - \left(\f{k_\perp\nu_\para}{6}\right)^2}
    - \ui \f{\nu_\para}{6} k_\perp^2 \ .
    \label{eq:fastwave-damping-main}
\en
In ideal MHD, fast magnetosonic waves propagate freely ($\gamma=0$) and are
non-dispersive ($\omega_0/k_\perp = \sqrt{\ca^2 + \va^2}$ is independent of
$k_\perp$) but Equation~\eqref{eq:fastwave-damping-main} shows that Braginskii
viscosity both damps and makes the waves dispersive. This also means that the
evolution of the velocity field is not perfectly out of phase with the density
and magnetic field evolution, i.e., Equation~\eqref{eq:fastwave-eig-v} has a
$\sin(\omega_0 t)$ contribution when $\gamma\neq0$. Using the full expressions
for the eigenmode, given by Equations~\eqref{eq:fastwave-eig-v} and
\eqref{eq:fastwave-eig-B}, is therefore important for obtaining good agreement
when comparing simulations and linear theory.

\begin{figure*}
    \begin{minipage}{\columnwidth}
        \includegraphics{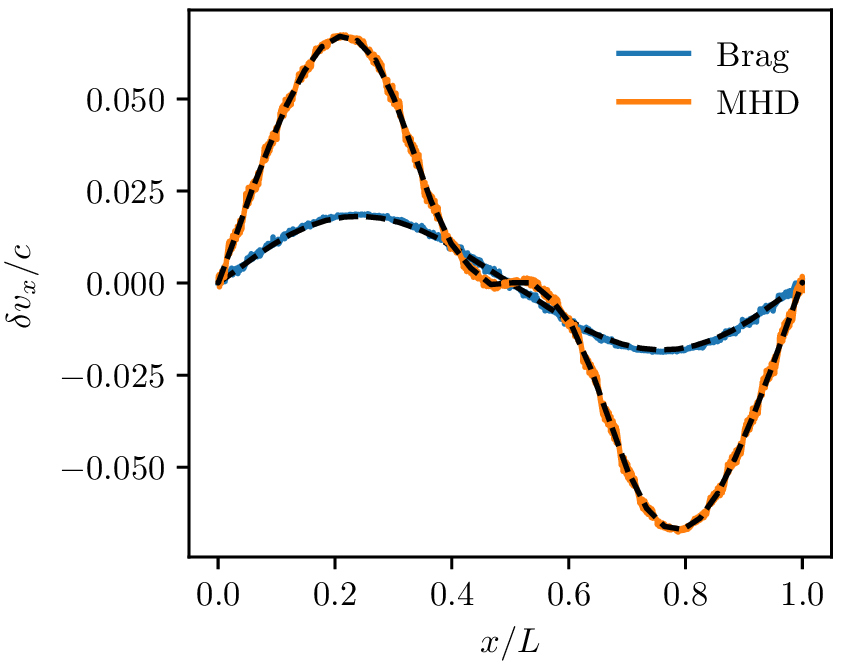}
        \caption{
        Fast magnetosonic waves in 3D. We compare the velocity profiles in
        ideal MHD and Braginskii MHD at $\cs t/L=1$. The ideal MHD simulation
        shows signs of nonlinearity at $x/L\approx 0.5$ while the profile in
        Braginskii viscosity is damped and retains a sinusoidal profile. The
        simulations use an irregular mesh created by adding, to each
        mesh-generating point, a random $\pm 0.2 \Dx$ offset in each
        direction. The dashed black lines are reference simulations using a
        regular mesh.
        }
        \label{fig:noisy-3D}
    \end{minipage}
    \hfill
    \begin{minipage}{\columnwidth}
        \includegraphics[trim = 0 15 0 0]{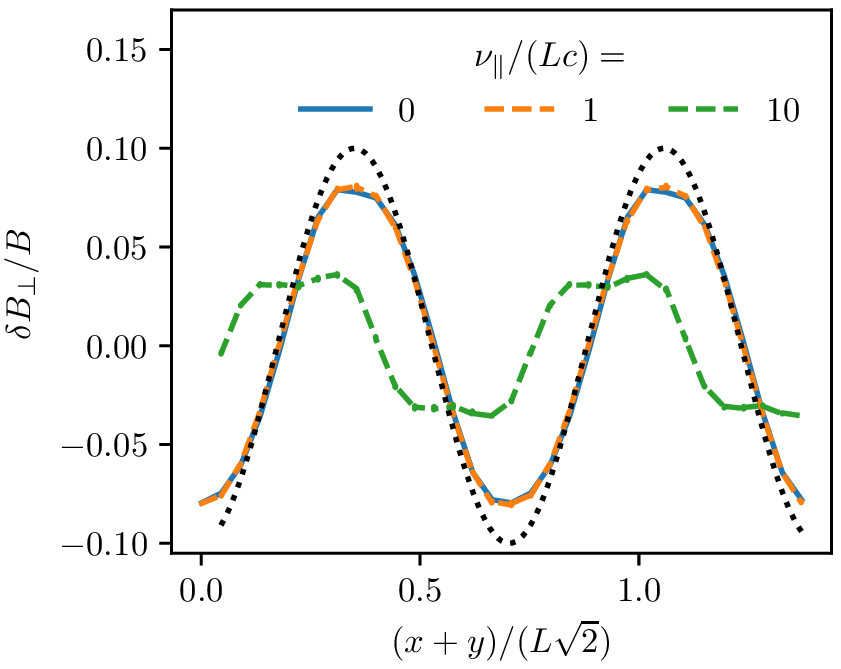}
        \caption{
        Damping of circularly polarized waves in a low resolution simulation
        after two wave periods ($\omega t= 4\pi$). The ideal MHD wave
        ($\nu_\para=0$) is damped solely due to inevitable numerical viscosity
        in \textsc{Arepo}'s MHD solver. The dotted black line shows the exact
        solution. A Braginskii simulation, which has $\nu_\para/(L\cs)=1$,
        which is $10^3$ times larger than the estimated numerical viscosity,
        suffers no additional damping. Only at $\nu_\para/(L\cs)=10$, do we
        find an additional erroneous damping of the wave. The extra numerical
        viscosity included by Braginskii viscosity is thus less than $10^{-3}
        \nu_\para$.
        }
        \label{fig:circ-wave}
    \end{minipage}
\end{figure*}

As an illustration of fast waves with Braginskii viscosity, we show the
evolution of the amplitude of the density perturbation for a simulation with
$\nu_\para/(L\cs)=0.05$ in the first panel of Fig.~\ref{fig:fast-waves}. We can
estimate the decay rate, $\gamma$, by fitting an exponential decay function to
the peaks in amplitude and the real frequency, $\omega_0$, by measuring the
distance between them. The agreement with the theoretical prediction for the
decay rate, which is given by $\gamma = \nu_\para k_\perp^2/6$ as long as
$k_\perp \nu_\para < 6 \sqrt{\ca^2 + \va^2}$, is excellent.

More quantitatively, we perform two parameter studies of damping of fast
magnetosonic waves. Firstly, we vary the value of the viscosity coefficient,
$\nu_\para$, at a fixed numerical resolution of $32\times 32$. We calculate
the damping rate and frequency for each simulation and show the result in the
middle panel of Fig.~\ref{fig:fast-waves}. The agreement is excellent except
for the data point with $\nu_\para/(L\cs)=10^{-3}$ where the damping rate is
higher in the simulation than it should be according to the linear theory.
This is due to numerical dissipation at the rather low numerical resolution
employed (see also \citealt{Par12} who reached the same conclusion for
\textsc{Athena} simulations).

The theoretical curves in the middle panel of
Fig.~\ref{fig:fast-waves} are computed from
Equation~\eqref{eq:fastwave-damping-main} which remains valid even in the
regime of strong damping and dispersion of fast waves (which occurs when
$k_\perp \nu_\para$ is high, i.e., when $k_\perp \nu_\para$ is not much
smaller than $\sqrt{\ca^2 + \va^2}$).
We are therefore able to  extend the results of
\citet{Par12} and test \textsc{Arepo} even in this regime.\footnote{\citet{Par12}
estimated the damping rate as
\be
    \gamma = \f{\nu_\para k^2}{6} \left[(\vec{\hat{k}}\bcdot \vec{\hat{\varv}})
    - 3 (\b \bcdot \vec{\hat{k}})(\b \bcdot \vec{\hat{\varv}})
    \right]^2 \ ,
\en
where $\vec{\hat{k}} = \vec{k}/k$ and  $\vec{\hat{\varv}}= \vec{\varv}/\varv$
but found that the damping rates in their simulations were lower than this
estimate at high $\nu_\para$. A solution of our
Equation~\eqref{eq:disp-sonic-waves} could in principle be used to understand
their simulations which considered a general $\vec{k}$ and both slow and fast
waves. } We find that \textsc{Arepo} is able to well describe the change in
real frequency and the purely exponential decay which occurs to the right of
the dashed vertical line in the middle panel of Fig.~\eqref{fig:fast-waves}.
These simulations were performed with the RKL2 super-time-stepping
method in order to expedite the process. This was particularly useful for the
two simulations with $\nu_\para/(L \cs)=5$ where we took $s_\mathrm{max}=17$.

Secondly, we vary the numerical resolution at a fixed value of the viscosity
coefficient, $\nu_\para/(L \cs) = 10^{-2}$. We then measure the relative error
in the measured damping rate,
$(\gamma_\mathrm{s}-\gamma_\mathrm{t})/\gamma_\mathrm{t}$, where
$\gamma_\mathrm{s}$ is the damping rate in the simulation and
$\gamma_\mathrm{t}$ is the theoretical damping rate. This error is shown as a
function of resolution $N=N_x=N_y$ in Fig.~\ref{fig:fast-waves}. We find third
order convergence, in agreement with \citet{Par12}.

As a final test, we consider fast magnetosonic waves on an irregular, random
mesh in 3D, with a significant amplitude ($A=0.1$) and
$\nu_\para/(L\cs)=0.05$.  We orient $\vec{k}=k\ex$ and construct a regular 3D
grid by interlacing two standard Cartesian 3D grids. This gives a grid with
$2N^3$ points where $N=16$. The noisy grid is then constructed by displacing
every mesh-generating point in all three directions by a random offset drawn
from the interval $[-0.2, 0.2]L/N$.

We perform four simulations, i.e., MHD and Braginskii MHD on both the regular
(for reference) and the irregular grid. The resulting velocity profile at $\cs
t/L=1$ is shown in Fig.~\ref{fig:noisy-3D}. The random mesh leads to noise in
the simulation results but overall, the solutions obtained with the random mesh
closely follows the results obtained with the regular mesh.

We also note that the significant initial amplitude of the wave ($A=0.1$) leads
to the development of small scale variation in the velocity profile in the
ideal MHD simulation only. In the Braginskii MHD such variations are damped on
an even shorter time scale than the damping of the large scale motion
(because the damping rate is inversely proportional to the length scale of
the variation, squared).

%-----------------------------------------------------------------------------
\subsection{Circularly polarized Alfv\'{e}n waves}
\label{sec:circ_waves}

Circularly polarized Alfv\'{e}n waves are not modified by Braginskii viscosity
because the associated pressure anisotropy is exactly zero. The magnetic
field strength and density is constant in time for such waves, which directly
follows from Equation~\eqref{eq:p-aniso}. This property should be respected
by the numerical implementation of Braginskii viscosity in Arepo. In
order to ensure this, we perform simulations of circularly polarized
Alfv\'{e}n waves. Any damping of their amplitude will be due to inherent
numerical viscosity in the main MHD solver of Arepo or due to interpolation
errors in the Braginskii viscosity implementation.

We initialize a traveling, circularly polarized Alfv\'{e}n wave by setting a
background magnetic field $\vec{B} = B(\ex + \ey)/\sqrt{2}$ and a perturbation
of the form
\be
    \f{\delta \vec{B}}{B} = A
    \left(
    \cos(\vec{k} \bcdot \vec{r}-\omega t)\f{\ey - \ex}{\sqrt{2}}
    - \sin(\vec{k} \bcdot \vec{r}-\omega t)\ez
    \right)  \ ,
\en
and
\be
    \delta \vec{\varv} = - \f{\omega}{k} \f{\delta \vec{B}}{B} \ .
\en
Here the wavevector is oriented along the background magnetic field,
$\vec{k} = k_\para \b$ with $k_\para = 2\sqrt{2}\pi/L$ and the frequency of
the Alfv\'{e}n wave is $\omega = k_\para \va$.

The simulation domain is a cubic box of size $L$ with $2N^3$ mesh-generating
points where $N=16$ and we let the wave propagate for two wave periods,
$\omega t= 4\pi$. We compare a simulation with a large viscosity coefficient,
$\nu_\para/(L\cs)=1$ (using RKL2 and $s_\mathrm{max}=17$), with a
reference simulation where Braginskii viscosity
is not included. Due to the low grid resolution, the amplitude of the wave
decays during its evolution in both simulations, see Fig.~\ref{fig:circ-wave}.
For the ideal MHD reference simulation, the decay is purely due to numerical
viscosity. By measuring the exponential decay rate, we find that
$\gamma_\mathrm{num}/(L\cs)\approx0.043$. We can estimate the corresponding
numerical viscosity coefficient, by assuming that the numerical viscosity can
be described as an isotropic viscosity of the form
$-\nu_\mathrm{num}\nabla^2 \vec{\varv}$ such that
$\gamma_\mathrm{num}=k^2\nu_\mathrm{num}$. This estimate gives
$\nu_\mathrm{num}/(L\cs)\approx 10^{-3}$. The Braginskii viscosity simulation
with $\nu_\para/(L\cs)=1$ does not show increased decay and this means that
the extra numerical viscosity added by the Braginskii viscosity is less than
$\nu_\mathrm{num}$.

Only at an explicit Braginskii viscosity coefficient of $\nu_\para/(L\cs)=10$
do we see a significant deviation from the reference simulation, see
Fig.~\ref{fig:circ-wave}. Note that this value of $\nu_\para$ is extremely
large and would lead to a decay rate of
$\gamma_\mathrm{iso}=k^2\nu_\mathrm{iso}\approx 400$ if it was added as
isotropic viscosity. The explicit decay rate would thus be $\sim 10^4$
times larger than the numerical decay rate instead of just a few times larger
as in Fig.~\ref{fig:circ-wave}. From the simulations we can conclude that the
extra numerical viscosity added by the Braginskii viscosity module is less
than $10^{-3}\nu_\para$. Similar encouraging findings have previously been
found for the Athena MHD code \citep[see the Appendix in][]{Par12}.

%-----------------------------------------------------------------------------
\subsection{Linearly polarized Alfv\'{e}n waves}
\label{sec:firehose}

In contrast to circularly polarized Alfv\'{e}n waves, linearly polarized
Alfv\'{e}n waves have an associated pressure anisotropy. Quite recently, it
was realized that this can prevent large amplitude linearly polarized
Alfv\'{e}n waves from propagating in a weakly collisional, high-$\beta$ plasma
because the oscillation is interrupted by the firehose instability
\citep{Squire2016}.

Extensive modeling of this phenomenon, using both a collisionless Landau fluid
closure and weakly-collisional Braginskii MHD, has been presented in
\citet{Squire2017}. The consequences for the magneto-rotational instability
(MRI) have also been investigated \citep{Squire2018} and the interruption was
studied using hybrid kinetic simulations in \citet{Squire2017b}.

\begin{figure*}
    \includegraphics[trim = 0 25 0 10]{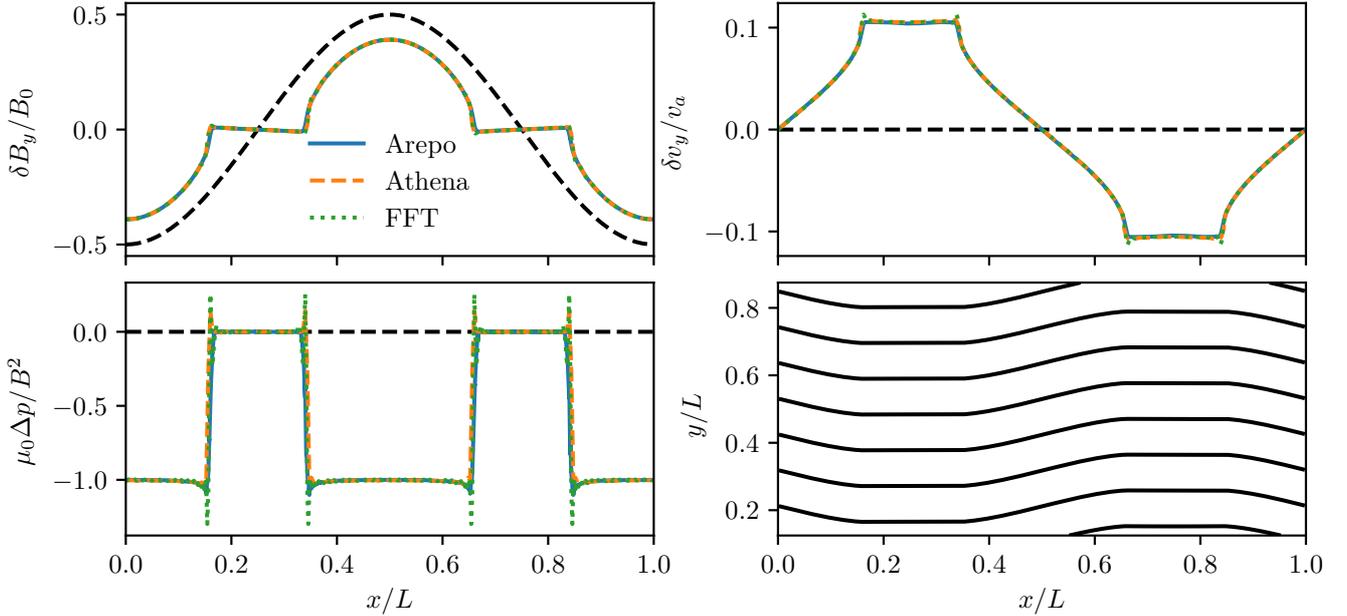}
    % trim={<left> <lower> <right> <upper>}
    \caption{
    Interruption of a standing, linearly polarized Alfv\'{e}n wave by the
    firehose instability. We compare simulations with \textsc{Arepo},
    \textsc{Athena} and a calculation of a simplified set of 1D equations
    which we solve using the fast Fourier transform (FFT). Dashed black lines
    show the initial conditions and simulations are shown at $\va t/L = 0.30$.
    Upper left: Magnetic field fluctuation. Upper right:
    Velocity fluctuation. Lower left: Pressure anisotropy normalized to
    $B^2/\mu_0$. Firehose instability limit is at $-1$ and mirror instability
    limit is at $0.5$. Lower right: Magnetic field lines in the plane
    of the simulation.
    }
    \label{fig:lin-polarized}
\end{figure*}

Here we compare an \textsc{Arepo} simulation with two reference solutions.
These are found by using the MHD code \textsc{Athena}
\citep{stone_athena:_2008} as well as a solution obtained by numerically
integrating a simplified set of 1D equations. Details of the latter
calculation, which is performed with spectral accuracy in space and a fourth
order accurate Runge-Kutta-Legendre update in time, can be found in
Appendix~\ref{sec:fft-calc}.

We initialize the simulations at $t=0$
with a magnetic field $\vec{B} = B_0 \ex$ and a perturbation of the form
\be
    \f{\delta \vec{B}}{B_0} = -A \cos\left(k x\right) \ey \ .
\en
where $k = 2\pi/L$ and the perturbation amplitude is $A=1/2$. This initial
profile is shown with a dashed black line in the upper left panel of
Fig.~\ref{fig:lin-polarized}. The initial magnetic field strength is such that
$\beta_0=10^4$ and the initial velocity is zero (shown in the upper right
panel of Fig.~\ref{fig:lin-polarized}).

In the absence of Braginskii viscosity, the resulting wave motion is a
linearly polarized Alfv\'{e}n wave with
\begin{align}
    \f{\delta \vec{B}}{B_0} &= -A \cos\left(k x\right) \cos(\omega t) \ey \ ,
    \label{eq:lin-pol-B}
    \\
    \f{\vec{\varv}}{\va} &= A \sin\left(k x\right) \sin(\omega t) \ey \ ,
    \label{eq:lin-pol-v}
\end{align}
where $\omega = k \va$ is the frequency of the wave.

When Braginskii viscosity is included in the simulation, this motion is
severely modified \citep{Squire2016,Squire2017}. This is evident in
Fig.~\ref{fig:lin-polarized} where both $\delta B_y$ and $\delta \varv_y$
have acquired highly non-sinusoidal shapes. The reason for this disruption of
the wave is a self-generated pressure anisotropy which can trigger the
firehose instability if the circumstances are right
\citep[see below, ][]{Squire2016}.

Unlike the circularly polarized Alfv\'{e}n wave, which has a constant in time
magnetic field strength, the linearly polarized Alfv\'{e}n wave given by
Equation~\eqref{eq:lin-pol-B} and \eqref{eq:lin-pol-v} has a magnetic field
strength that varies as
\be
    B(t) = B_0 \sqrt{1 + A^2 \cos^2(k x)\cos^2(\omega t)} \ .
\en
This decaying magnetic field strength leads to a pressure anisotropy given by
\be
    \Delta p(x, t) = 3\rho \nu_\para \der{\ln B}{t}
    =
    - \f{3\rho \nu_\para A^2 \omega \cos^2(k x) \sin(2 \omega t)}
    {2 \left(1 + A^2 \cos^2(k x) \cos^2(\omega t)\right)}
    \ ,
\en
which eventually triggers the firehose instability if there are regions where
Equation~\eqref{eq:firehose} is fulfilled, i.e., if the following inequality
is satisfied
\be
    \mu_0\f{\Delta p}{B_0^2} =
    - \f{3\nu_\para A^2 \omega \cos^2(k x) \sin(2 \omega t)}
    {2 \va^2 \left[1 + A^2 \cos^2(k x) \cos^2(\omega t)\right]^2}
    < - 1\ .
    \label{eq:pa-squire-test}
\en
From Equation~\eqref{eq:pa-squire-test} we observe that the pressure anisotropy will
attain its extrema at $x/L=0$ and $x/L=1/2$ while the pressure anisotropy at
$x/L=1/4$ and $x/L=3/4$ remains zero.

As in \citet{Squire2017}, we can estimate whether the firehose instability
will be triggered at the extrema of the pressure anisotropy by setting $\cos^2
(kx) = 1$ in Equation~\eqref{eq:pa-squire-test}. If $A$ is reasonably small,
we can furthermore ignore the squared parenthesis in the denominator. We then
find that
\be
    \mu_0\f{\Delta p}{B^2} \approx
    - \f{3\nu_\para A^2 \omega \sin(2 \omega t)}
    {2 \va^2}
    < - 1\ .
    \label{eq:squire-ampl-with-sin}
\en
Since the maximum of $\sin(2 \omega t)$ is 1, the criterion for interruption
of the linearly polarized Alfv\'{e}n wave becomes\footnote{We note that
$\nu_\mathrm{c}$ in the notation of \citet{Squire2017} is related to
$\nu_\para$ by $\nu_\mathrm{c} \nu_\para = \cs^2/3$. Using also that
$\beta = 2\cs^2/\va^2$ it is easy to show that
Equation~\eqref{eq:squire-ampl} agrees with equation 30 in
\citet{Squire2017}.}
\be
    A \gtrsim
    \sqrt{\f{2 \va^2}{3\nu_\para \omega}}
    \label{eq:squire-ampl}
    \ .
\en
As in \citet{Squire2016}, we choose the viscosity coefficient
such that the RHS of Equation~\eqref{eq:squire-ampl} is $1/5$, i.e.,
\be
    \nu_\para = \f{50 \va^2}{3 \omega}
    \ ,
\en
which has the numerical value $\nu_\para/(L\cs)= 3.75 \times 10^{-2}$.
With a perturbation amplitude $A=1/2$, we thus expect the wave to be
interrupted by the firehose instability.

This prediction is fulfilled by the simulations, where interruption of the
wave leads to large regions in which the pressure anisotropy divided by the
magnetic field strength squared, i.e., $\mu_0\Delta p/B^2$, is constant. Its
value is either zero or pinned to the firehose instability threshold,
$\mu_0\Delta p/B^2=-1$ (see the lower left panel of
Fig.~\ref{fig:lin-polarized}). In both types of regions, the wave evolves more
slowly than it would have in the absence of Braginskii viscosity. This happens
because the restoring force of the wave is the sum of the Lorentz force and
the pressure anisotropy. The regions where $\mu_0\Delta p/B^2$ is zero also
have zero magnetic field fluctuation, as can be seen by comparing the lower
and upper left panels of Fig.~\ref{fig:lin-polarized}. There is thus no
restoring force for the wave. In the regions where the magnetic field
fluctuation is non-zero, the restoring force is canceled by the pressure
anisotropy, $\Delta p + B^2/\mu_0=0$. In real space, the wave evolution leads
to a zigzag pattern in the magnetic field structure, with segments of straight
field lines that are connected with curved field segments in which the
firehose instability is at the instability threshold (see the lower right
panel of Fig.~\ref{fig:lin-polarized}).

A more detailed account of the test problem described in this section as well
as the behavior found for traveling waves
can be found in \citet{Squire2016} and \citet{Squire2017}.

%-----------------------------------------------------------------------------
\subsection{The viscous Kelvin-Helmholtz instability}
\label{sec:KHI}

\begin{figure*}
    % trim={<left> <lower> <right> <upper>}
    \includegraphics[trim= 0 15 0 0]{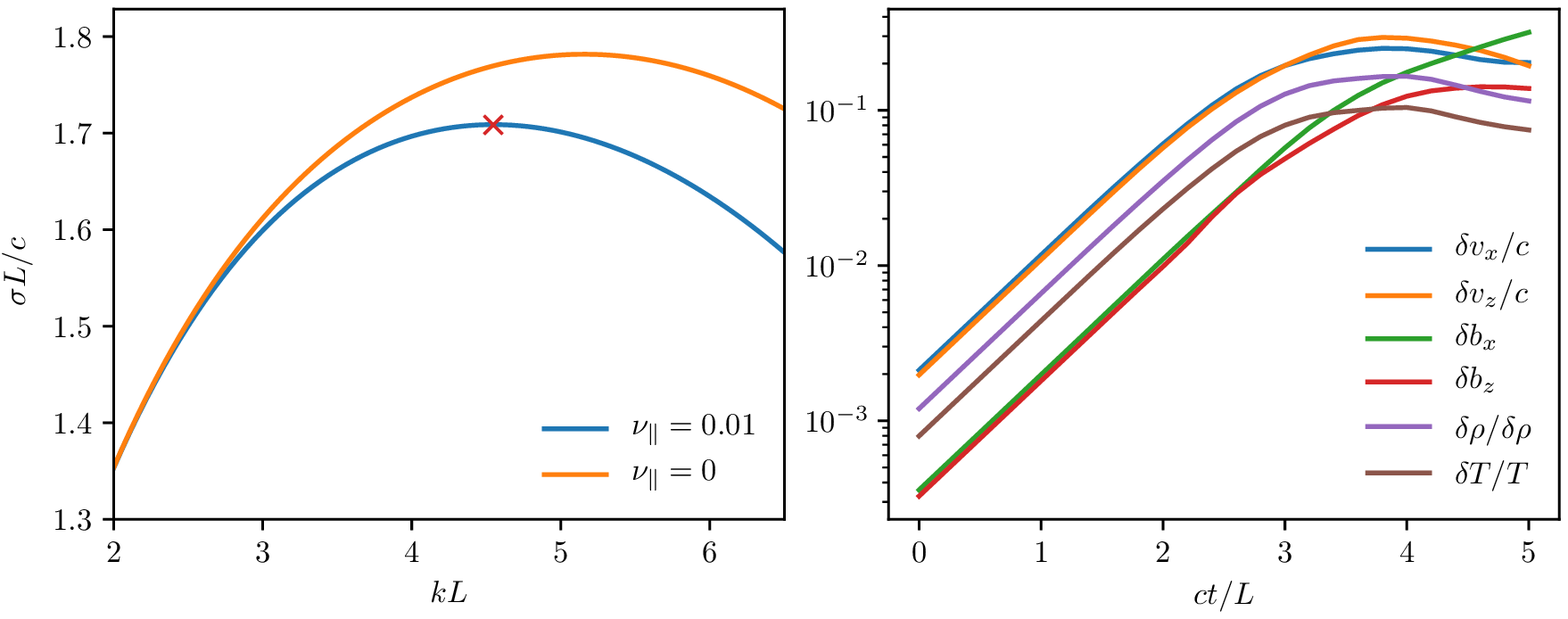}
    \caption{
    Left:
    growth rates as a function of wavenumber, $k L$, for $\nu_\para=0$ and
    $\nu_\para/(L\cs)=0.01$ obtained with \textsc{Psecas} (solid lines)
    and the growth rate found in the \textsc{Arepo} simulation without
    pressure anisotropy limiters (red cross).
    Braginskii viscosity inhibits
    growth of the KHI and moves the maximally unstable mode to longer
    wavelengths. Right: exponential growth of deviations from the
    background equilibrium as measured in the \textsc{Arepo} simulation.
    }
    \label{fig:kh-linear-theory}
    %second-figure
    \includegraphics[trim= 0 20 0 0]{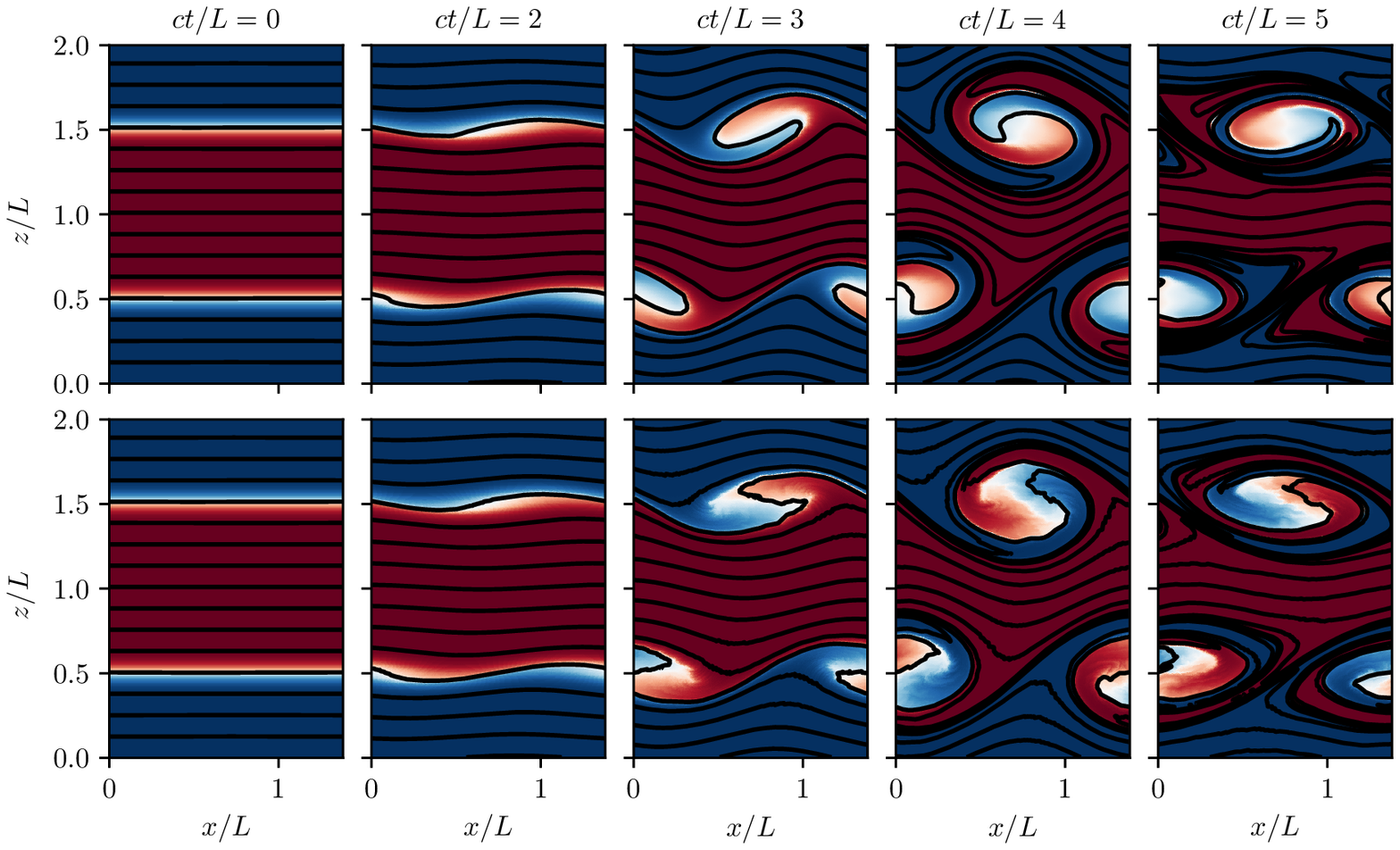}
    \caption{Evolution of passive scalar field (false color image)
    and magnetic field lines (black solid lines) as a function of time.
    The simulation in the upper panels has the pressure anisotropy limited to
    lie inside the microscale instability thresholds
    (Equation~\ref{eq:fire_and_mirror}) while the pressure anisotropy is
    allowed to freely evolve in the simulation shown in the lower panels.}
    \label{fig:kh-simulation-snapshots}
\end{figure*}

We study the KHI in a weakly collisional, magnetized plasma by including
Braginskii viscosity in our analysis. This test is of particular relevance for
cold fronts in galaxy clusters (see the discussion in
Section~\ref{sec:intro}). We consider a 2D periodic domain where the
$x$-component of the velocity has a smooth profile given by
$\varv_x(z) = \varv(z) \ex$ with
\begin{align}
    \varv(z) = \varv_0 \left[\tanh\left(\f{z-z_1}{a}\right) -
                       \tanh\left(\f{z-z_2}{a}\right)\right] \ ,
                       \label{eq:smooth-v}
\end{align}
where $z_1=0.5 L$, $z_2=1.5 L$, the smoothing
parameter is $a = 0.05 L$ and $2L$ is the extent of the domain in the
$z$-direction. Here the flow
speed is $\varv_0/\cs=1$ where $\cs$ is the constant, isothermal sound speed
of the background. Finally, we include a dynamically weak background magnetic
field, $\vec{B} = B\ex$, with $B$ such that $\beta=10^3$.

Using a smooth profile such as Equation~\eqref{eq:smooth-v}, instead of a
discontinuous velocity profile, is essential for obtaining converged results
for the KHI \citep{McNally2012,Lecoanet2016}. The linear theory, however,
becomes analytically intractable due to the $z$-dependence of the background
velocity shear \citep{Berlok2019}. It is nevertheless possible to numerically
calculate the growth rate of the KHI for smooth shear profiles by linearizing
Equations~\eqref{eq:rho}-\eqref{eq:ent} and using a pseudo-spectral method to
generate a matrix eigenvalue problem. This procedure is described in detail in
\citet{Berlok2019} where it is applied to the KHI in various regimes, e.g., in
sub- and supersonic flows, with and without background density variation,
magnetic field and Braginskii viscosity. The linear calculations are partially
automated by using \textsc{Psecas}\footnote{Psecas (Pseudo-Spectral Eigenvalue
Calculator with an Automated Solver) is freely available online
\citep{Berlok2019}.} and the linear solutions can be used to compare with
nonlinear computer simulations. Such a verification test was developed in
\citet{Berlok2019} for $\nu_\para/(L\cs)=0.01$ (referred to as M$\nu$ in Table
2 therein) and will here be used to test Braginskii viscosity in
\textsc{Arepo}.

\begin{figure*}
    %first-figure
        \includegraphics[trim= 0 15 0 0]{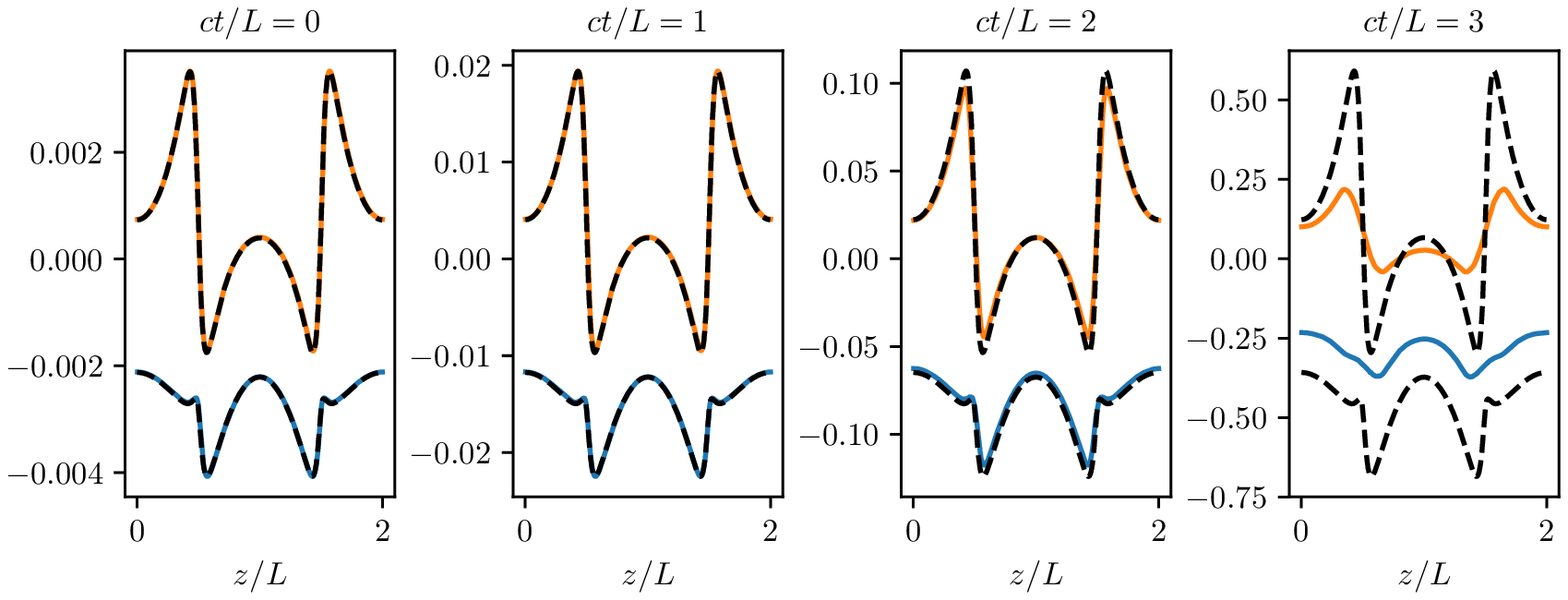}
        \caption{
        Growth of the perturbation to $\varv_z$ in the simulation (solid lines)
        compared to the exponential growth predicted by theory (dashed black
        lines). Blue and orange lines show the $z$-dependent amplitude of
        the sine perturbation, $S_k(z)$, and the cosine perturbation, $C_k(z)$.
        Nonlinear effects cause saturation of the instability and the linear
        theory therefore over-predicts the amplitude at $\cs t/L=3$. Note the
        change in scale on the vertical axis between panels.
        }
        \label{fig:kh-vz-mode-structure}
    %second-figure
    \includegraphics[trim= 0 20 0 0]{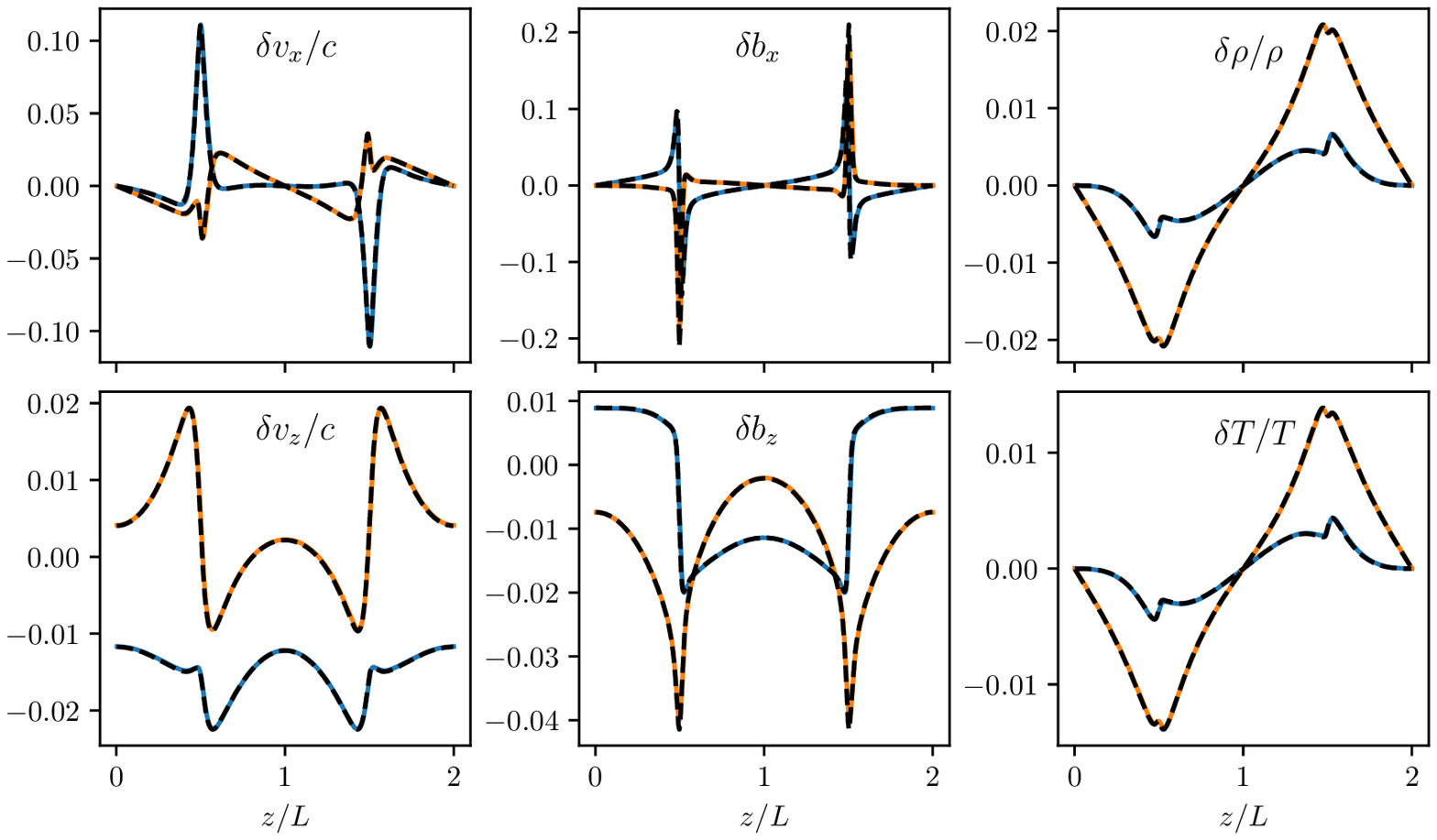}
    \caption{
    Eigenmode structure of the KHI with Braginskii viscosity at $\cs t/L = 1$.
    At this point in time the perturbations have grown by a factor of 5 and we
    still see agreement between simulation (blue and orange solid lines) and
    the linear theory (dashed black lines).
    }
    \label{fig:kh-all-modes}
\end{figure*}

The growth rates obtained with \textsc{psecas} for the KHI with a Braginskii
viscosity coefficient $\nu_\para/(L\cs)=0.01$ are shown in the left panel of
Fig.~\ref{fig:kh-linear-theory} along with the inviscid solution for
comparison. As expected, the solution including Braginskii viscosity has a
slightly lower growth rate than the inviscid solution. Furthermore, the growth
rate maximum is moved to longer wavelengths (lower wavenumbers) because
viscosity acts most efficiently on short spatial scales. The maximum growth
rate is $\sigma_{\mathrm{max}}L/c=1.7087545$ and occurs at $k_{\mathrm{max}}
L = 4.5470431$ \citep{Berlok2019}.

We perform two simulations of the KHI using \textsc{Arepo} with
Braginskii viscosity enabled with $\nu_\para/(L\cs)=0.01$. One simulation
in which
the pressure anisotropy is limited to lie within the threshold for stability
of microscale instabilities (see Section~\ref{sec:firehose_and_mirror} for
details) and one in which the pressure anisotropy is allowed to freely evolve.
We set the extent of the domain in the
$x$-direction, $L_x$, to be such
that the fastest growing mode fits exactly inside the domain,
$L_x = 2\upi/k_{\mathrm{max}}$, and the instability is seeded by initializing
all quantities with the linear solution for the perturbations (see
\citet{Berlok2019} for details).
We use a moving mesh
with $354\times 512$ mesh-generating points and
the RKL2 method with $s_\mathrm{max}=11$
in order to
accelerate the simulation. The initial
amplitude of the perturbations is $10^{-2}$
but as the system evolves the perturbations grow exponentially until they
nonlinearly saturate (see right panel of Fig. \ref{fig:kh-linear-theory}). The
exponential growth closely matches the theoretical value (the fit to
$\delta \rho/\rho$ yields $\sigma L/c= 1.7084$ in the simulation without
$\Delta p$-limiters).\footnote{The simulation with $\Delta p$-limiters shows a slight
$\sim 2 \%$ deviation in the growth rate. A deviation is expected since the $\Delta p$-limiters were not included in the
theory of \citet{Berlok2019}.}

We show the evolution of the KHI in Fig.~\ref{fig:kh-simulation-snapshots}
with black solid lines indicating the initially straight magnetic field lines.
A passive scalar is shown in order to track the movement of the gas with red
(blue) indicating initially right-moving (left-moving) fluid. The nonlinear
evolution leads to the classical KHI rolls, and associated winding up of the
magnetic field lines, see Fig.~\ref{fig:kh-simulation-snapshots}.
We show the simulation with $\Delta p$-limiters
(upper row of panels) and the simulation without $\Delta p$-limiters
(lower row of panels)
where microscale are allowed to grow without bounds.
Microscale instabilities make the magnetic field become
more jagged which makes the simulation without $\Delta p$-limiters
(bottom row) appear less viscous than the one with $\Delta p$-limiters
(upper row) where the solution remains smooth.
However, the simulation without $\Delta p$-limiters is actually more viscous
because large pressure
anisotropies lead to increased damping of the velocity-field. This can be
seen from the less evolved stage of the KHI rolls in the lower row of
Fig.~\ref{fig:kh-simulation-snapshots} in comparison to
the upper row.

As mentioned above, the linear solution is seeded exactly with a single mode
in the $x$-direction,
where, e.g., the $\dvz$-component of the solution can be written as
\be
    \dvz(x, z, t) = c \left[C_k(z)\cos(k x) + S_k(z)\sin(k x)\right] \ .
\en
Here $C_k(z)$ and $S_k(z)$ are $z$-dependent Fourier-amplitudes that grow
exponentially in time. In order to assess how well \textsc{Arepo} retains the
correct shape of the linear solution, we calculate $C_k(z)$ and $S_k(z)$ from
simulation snapshots at four different times and compare with the linear
theory in Fig.~\ref{fig:kh-vz-mode-structure}. We find an exact match at $t=0$
(simply due to the seeding), and a very good match after amplification by a
factor of $\sim 5$ at $\cs t/L = 1$. Nonlinear interactions start to play a
significant role at $\cs t/L = 2$ and cause the linear solution to
over-predict the amplitudes by a factor of two at $\cs t/L = 3$.

Finally, we show the spatial profiles for all the components of the system
at $\cs t/L = 1$ in Fig.~\ref{fig:kh-all-modes}. This figure shows an
excellent (but at this numerical resolution not perfect) match between linear
theory and \textsc{Arepo} simulation.
The very sharp variations with $z$ in the profiles (in particular for
$\delta b_x = \delta B_x/B$) explain why a high grid resolution is required to
accurately simulate the linear evolution of the KHI with magnetic fields and
Braginskii viscosity.

\subsection{Verification of RKL2 super-time-stepping method}

The RKL2 method has been employed in several of the tests presented in
the previous subsections. In this section we briefly verify that
the method is second order accurate in time and that it gives the
expected reduction in computational cost.

\begin{figure}
\includegraphics[trim= 0 25 0 0]{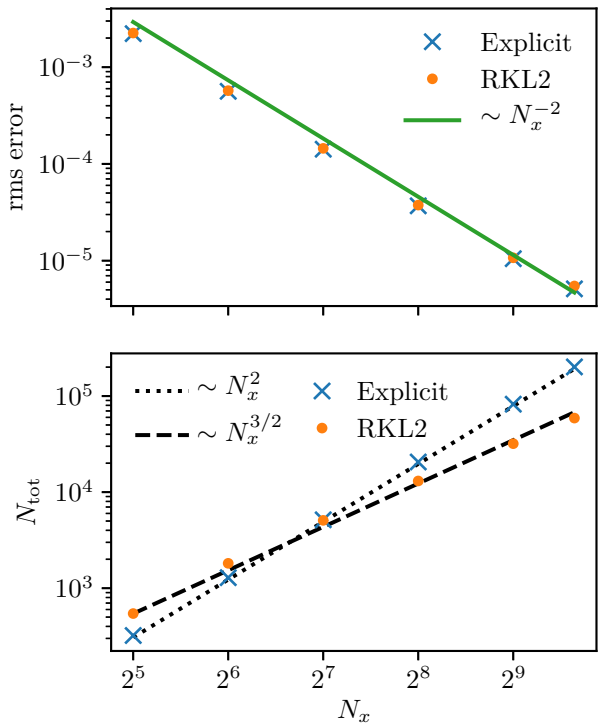}
\caption{Verification of RKL2 super-time-stepping. Upper:
The RMS difference between simulation and analytic theory
at $\cs t/L =1$ as a function of grid resolution. Lower:
The total number of computational steps needed to reach $\cs t/L =1$
as a function of grid resolution.
}
\label{fig:sts_check}
\end{figure}

We consider the anisotropic diffusion test presented in
Section~\ref{sec:decay2} but increase the value of $\nu_\para$ by
a factor of 10. We perform six simulations with increasing spatial
resolution. Following the procedure used in \citet{Vaidya2017},
we set the ratio of super time step to grid spacing, $\tau/\Dx$,
to be constant. The resulting number of STS stages then increase in
increments of 2 from $s=3$ at the lowest resolution to $s=13$ at the highest.
We calculate the
root-mean-square (RMS) differences between analytic theory and
internal energy, $\varepsilon$, in the simulations
(at $\cs t/L =1$) and find second order convergence, see
the upper panel in Fig.~\ref{fig:sts_check}. We show the corresponding
computational cost, calculated as the total number of computational steps
required to reach $\cs t/L =1$ (i.e. $N_\mathrm{tot} = t/\tau s$), in the
lower panel of Fig.~\ref{fig:sts_check}. We find, in agreement with
expectations \citep{Vaidya2017}, that the number of
total steps scales as $N_x^{3/2}$. This is a factor $\sqrt{N_x}$ less
than the scaling for the number of steps needed for a standard explicit update.

In general, the different scalings of the explicit Braginskii time step,
$\Dt\propto \Dx^2$, and the MHD time step, $\Dt_\mathrm{MHD}\propto \Dx$,
makes the number of required stages scale as $s\propto 1/\sqrt{\Dx}$
(see Equations~\ref{eq:dt-brag},
\ref{eq:dt-MHD} and \ref{eq:sts-tau}). The total number of computational
steps then scales as $N_\mathrm{tot} \propto s/\Dt_\mathrm{MHD}
\propto \Dx^{-3/2}$ when using RKL2 and as $N_\mathrm{tot} \propto \Dx^{-2}$
when using an explicit update.\footnote{In practice, there is a maximum number of stages,
$s_\mathrm{max}$, above which operator splitting of diffusion and MHD
and/or the accuracy of RKL2 breaks down. The best
value for $s_\mathrm{max}$ is likely to be problem dependent and we are
not aware of a systematic approach to determining it. One potential idea,
see the Appendix in
\citet{Hopkins2017}, is to define a signal speed for the diffusive flux,
$\varv_{\mathrm{diff}}$, and use that to calculate the maximum
super-time-step (and thereby $s_\mathrm{max}$) as
$\tau = C \Dx/\varv_{\mathrm{diff}}$ at the beginning of each
super-time-step.
}
We conclude that the RKL2 method can accelerate the Braginskii viscosity update
while retaining second order accuracy.

%-----------------------------------------------------------------------------
%-----------------------------------------------------------------------------
\section{Conclusion and future prospects}
\label{sec:conclusion}

We have presented a new scheme for modeling Braginskii viscosity on an
unstructured, moving mesh. This scheme has been implemented in the MHD code
\textsc{Arepo} \citep{Springel2010,Pakmor2011} which already had the
capability of modeling anisotropic heat conduction \citep{Kannan2016}. With
methods for describing both anisotropic heat conduction and viscosity in
place, \textsc{Arepo} is now capable of describing weakly collisional plasmas
within the Braginskii MHD framework.

The Braginskii viscosity implementation has been thoroughly tested against
analytical solutions (some newly derived) and independent numerical solutions.
The examples show that the inclusion of Braginskii viscosity can change,
sometimes fundamentally, the system evolution. The first example considered
the decay of a velocity profile aligned with the magnetic field. Secondly, we
rotated the magnetic field as an illustration of how Braginskii viscosity can
mix velocity components. We were able to provide additional insight into this
example, originally presented by  \citet{Hopkins2017}, by deriving and
comparing with the analytical solution and also analyzing the associated
viscous heating.

Thirdly, we tested the ability of our code to propagate the MHD wave family
with the correct modified behavior due to Braginskii viscosity. We found, in
agreement with linear theory, that fast magnetosonic waves are damped by
dissipation of their self-generated pressure anisotropy. We also verified that
circularly polarized Alfv\'{e}n waves propagate without damping\footnote{Even
circularly polarized waves are in fact damped by numerical dissipation. What
we have verified is that the damping rate is not increased by including
explicit Braginskii viscosity.}, as they should because they have a constant
magnetic field strength and density and, consequently, zero pressure
anisotropy  (see Equation~\ref{eq:p-aniso}).  Linearly polarized Alfv\'{e}n
waves, on the other hand, have a time-varying magnetic field strength with an
associated pressure-anisotropy that can lead to non-trivial evolution of the
wave structure. This was realized only recently by \citet{Squire2016} who
found that triggering of the firehose instability happens in a high-$\beta$
plasma whenever the wave-amplitude exceeds a certain limit (given in
Equation~\ref{eq:squire-ampl}). The Alfv\'{e}n wave plays a special role
in magnetized turbulence theory \citep{Goldreich1995}, and
Alfv\'{e}n wave interruption therefore modifies
turbulence properties in weakly collisional plasmas
\citep{Squire2017,Squire2019}. It is therefore promising, that our
implementation in \textsc{Arepo} is able to reproduce Alfv\'{e}n wave
interruption and that we find agreement with both the well-tested MHD code
\textsc{Athena} and an independent numerical solution obtained with spectral
methods.

We concluded the tests by performing a simulation of the KHI in a weakly
collisional, magnetized plasma. In this test, we initialized the simulation
with a highly precise linear solution obtained in \citet{Berlok2019}. The
\textsc{Arepo} simulation accurately follows the linear solution until the
amplitude of the disturbance has grown large enough for nonlinear interactions
to modify the evolution and lead to saturation. We find that Braginskii
viscosity inhibits the growth rate of the KHI in agreement with theory
\citep{Suzuki2013,Berlok2019} and previous numerical studies of cold fronts in
galaxy clusters \citep{ZuHone2015}.

Our implementation employs 2nd order accurate (in time) Runge-Kutta-Legendre
super-time-stepping (RKL2, \citealt{Meyer2014}) which is used to alleviate the
severe time step constraint associated with diffusion operators in
simulations. This method has previously been used for anisotropic heat
conduction on a Cartesian static grid but it is, to our knowledge, the first
use of the method for Braginskii viscosity and the first use of this method on
an unstructured, moving grid. The use of RKL2 will allow us to perform
Braginskii MHD simulations of AGN-inflated bubbles and sloshing cold fronts in
galaxy clusters without the computational cost becoming prohibitively
expensive.

%-----------------------------------------------------------------------------
%-----------------------------------------------------------------------------
\section*{Acknowledgments} It is a pleasure to thank Volker Springel for
the use of \textsc{Arepo} and the referee for useful recommendations.
We thank Jonathan Squire for sharing his numerical solutions of
Alfv\'{e}n wave interruption and Jonathan Squire, Matthew Kunz and Jim
Stone for an interesting discussion during the KITP program
\emph{Multiscale Phenomena in Plasma Astrophysics}.
T.B. and C.P. acknowledges support by the European
Research Council under ERC-CoG grant CRAGSMAN-646955. This research was
supported in part by the National Science Foundation under Grant No. NSF
PHY-1748958.

\bibliographystyle{mnras}
\bibliography{references}

\appendix

%-----------------------------------------------------------------------------
%-----------------------------------------------------------------------------
\section{Coordinate transformations}
\label{sec:coord-trans}

Any vector, for instance the velocity vector $\vec{\varv} = \varv_x \ex +
\varv_y \ey + \varv_z \ez$, can be expressed in terms of the new coordinate
vectors $\un$, $\um$ and $\up$ as
\be
    \vec{\varv} = \left(\un \bcdot \vec{\varv}\right) \un +
             \left(\um \bcdot \vec{\varv}\right) \um +
             \left(\up \bcdot \vec{\varv}\right) \up =
              \varv_n \un +
              \varv_m \um +
              \varv_p \up
             \ .
\en
Similarly the vector
\be
\vec{\varv} =
              \varv_n \un +
              \varv_m \um +
              \varv_p \up
             \ ,
\en
in the local coordinate system, is given by
\be
    \vec{\varv} =
    \left(\ex \bcdot \vec{\varv}\right) \ex +
                 \left(\ey \bcdot \vec{\varv}\right) \ey +
                 \left(\ez \bcdot \vec{\varv}\right) \ez =
                  \varv_x \ex +
                  \varv_y \ey +
                  \varv_z \ez
                 \ ,
\en
in the Cartesian coordinate system.
We can write out the transformation rules explicitly by defining the
components of the unit vectors $\un$, $\um$ and $\up$ in the Cartesian basis.
These are given by $\un = n_x \ex + n_y \ey + n_z \ez$, $\up = p_x \ex + p_y
\ey + p_z \ez$ and $\um = m_x \ex + m_y \ey + m_z \ez$. Transformations
between $(\varv_x, \varv_y, \varv_z)$ and $(\varv_n, \varv_m, \varv_p)$ are
then given by
\be
    \left(
    \begin{matrix}
    \varv_n \\ \varv_m \\ \varv_p
    \end{matrix}
    \right) = \mathsf{R} \bcdot
    \left(
    \begin{matrix}
    \varv_x \\ \varv_y \\ \varv_z
    \end{matrix}
    \right) \ ,
    \quad
    \left(
    \begin{matrix}
    \varv_x \\ \varv_y \\ \varv_z
    \end{matrix}
    \right) = \mathsf{R}^{T} \bcdot
    \left(
    \begin{matrix}
    \varv_n \\ \varv_m \\ \varv_p
    \end{matrix}
    \right) \ ,
\en
where the matrix $\mathsf{R}$ is given by
\be
    \mathsf{R} =
    \left(
    \begin{matrix}
    n_x & n_y & n_z \\
    m_x & m_y & m_z \\
    p_x & p_y & p_z
    \end{matrix}
    \right) \ .
\en

%-----------------------------------------------------------------------------
\subsection{Converting gradients to the local coordinate system}
\label{sec:convert-grad-to-loc}

According to Equation~\eqref{eq:p_aniso_loc}, we only require 5 of the nine
components of $\del \vec{\varv}$ in the local coordinate system. These are
given by
\begin{align}
\label{eq:dn_vn}
    \partial_n \varv_n &=
    \un \bcdot \del \left(\un \bcdot \vec{\varv} \right) = \nonumber \\ &
    \left(n_x \pder{}{x} + n_y \pder{}{y} + n_z \pder{}{z}\right)
    \left(n_x \varv_x + n_y \varv_y + n_z \varv_z\right) \ , \\
    \partial_m \varv_m &=
    \um \bcdot \del \left(\um \bcdot \vec{\varv} \right) = \nonumber \\ &
    \left(m_x \pder{}{x} + m_y \pder{}{y} + m_z \pder{}{z}\right)
    \left(m_x \varv_x + m_y \varv_y + m_z \varv_z\right) \ , \\
    \partial_n \varv_m &=
    \un \bcdot \del \left(\um \bcdot \vec{\varv} \right) = \nonumber \\ &
    \left(n_x \pder{}{x} + n_y \pder{}{y} + n_z \pder{}{z}\right)
    \left(m_x \varv_x + m_y \varv_y + m_z \varv_z\right) \ , \\
    \partial_m \varv_n &=
    \um \bcdot \del \left(\un \bcdot \vec{\varv} \right) = \nonumber \\ &
    \left(m_x \pder{}{x} + m_y \pder{}{y} + m_z \pder{}{z}\right)
    \left(n_x \varv_x + n_y \varv_y + n_z \varv_z\right) \ , \\
    \partial_p \varv_p &=
    \up \bcdot \del \left(\up \bcdot \vec{\varv} \right) = \nonumber \\ &
    \left(p_x \pder{}{x} + p_y \pder{}{y} + p_z \pder{}{z}\right)
    \left(p_x \varv_x + p_y \varv_y + p_z \varv_z\right) \ .
    \label{eq:dp_vp}
\end{align}

%-----------------------------------------------------------------------------
%-----------------------------------------------------------------------------
\section{Equations in the local coordinate system}
\label{app:local-equations}

We convert the momentum equations to the local coordinate system and find
using Equation~\eqref{eq:Pi} that
\be
\mathbf{\Pi} \bcdot \un = \Delta p \left(\b\b -\mathbf{1}/3\right)
    \bcdot \un =
    \Delta p \b b_n
    - \Delta p \un/3 \ .
\en
The terms needed in the momentum equations  are then
\begin{align}
    \left(\mathbf{\Pi} \bcdot \un\right) \bcdot \un &=
    \Delta p b_n b_n - \Delta p/3 \ , \\
    \left(\mathbf{\Pi} \bcdot \un\right) \bcdot \um &=
    \Delta p b_m b_n \ , \\
    \left(\mathbf{\Pi} \bcdot \un\right) \bcdot \up &=
    0 \ ,
\end{align}
and we observe that the last component of $\vec{\varv}$ does not evolve when
working in the local coordinate system.

We derive an equation for the energy flux $\mathbf{\Pi} \bcdot \un$ in the
local coordinate system. Using Equation~\eqref{eq:Pi} we obtain
\begin{align}
    \mathbf{\Pi} \bcdot \vec{\varv} = \Delta p \left(\b\b -\mathbf{1}/3\right)
    \bcdot \vec{\varv} =
    \Delta p \b \left(b_n \varv_n + b_m \varv_m\right)
    - \Delta p \vec{\varv}/3 \ ,
\end{align}
and the expression needed in the energy equation is thus
\be
    \left(\mathbf{\Pi} \bcdot \vec{\varv}\right)\bcdot \un
    = \Delta p \left(b_n b_n \varv_n + b_n b_m \varv_m - \varv_n/3 \right) \ .
\en
The pressure anisotropy given by Equation~\eqref{eq:p-aniso} can be expressed
in the local coordinate system by combining
\be
    \b\b\vec{:}\del \vec{\varv} =
    b_n^2 \partial_n \varv_n +
    b_m^2 \partial_m \varv_m +
    b_n b_m \left(\partial_m \varv_n + \partial_n \varv_m \right) \ ,
\en
and
\be
    \del \bcdot \vec{\varv} = \partial_n \varv_n + \partial_m \varv_m +
    \partial_p \varv_p .
\en
to give Equation~\eqref{eq:p_aniso_loc}.

%-----------------------------------------------------------------------------
%-----------------------------------------------------------------------------
\section{Analytical solutions to the diffusion equations}
\label{sec:diff-solutions}

We consider a plasma with a constant density and a uniform magnetic field. We
also assume that the viscosity coefficient is constant. This assumption
makes the velocity evolution independent from the internal energy evolution.
In 2D we work in a Cartesian coordinate system aligned with the magnetic field
by defining unit vectors parallel, $\para$, and perpendicular, $\perp$, to the
magnetic field.

In 3D we consider polar coordinates ($z, r, \phi$) corresponding to the
parallel, perpendicular and azimuthal direction. We assume for simplicity that
all fields are independent of the $\phi$ coordinate in 3D. The merit of this
coordinate system is that the equations simplify considerably. We have that
\be
    \b\b = \left(
    \begin{matrix}
    1 & 0 \\
    0 & 0
    \end{matrix}
    \right) \ ,
\en
\be
    \nabla \varv =
    \left(
    \begin{matrix}
    \partial_\para \varv_\para & \partial_\para \varv_\perp \\
    \partial_\perp \varv_\para & \partial_\perp \varv_\perp
    \end{matrix}
    \right) \ ,
\en
such that
\be
    \Delta p = \rho \nu_\para \left(2 \partial_\para \varv_\para -
                                      \partial_\perp \varv_\perp\right) \ .
    \label{eq:pa-in-parallel}
\en
In both the 2D and 3D case,
Equations~\eqref{eq:momentum-eq} and \eqref{eq:p-aniso} then become

\begin{align}
    \pder{\varv_\para}{t} &= \f{2\nu_\para}{3}
    \left(2\partial_\para^2\varv_\para -
    \partial_\para \partial_\perp \varv_\perp \right) \ ,
    \label{eq:v_coupled1}
    \\
    \pder{\varv_\perp}{t} &= -\f{\nu_\para}{3}
    \left(2\partial_\perp \partial_\para
    \varv_\para - \partial_\perp^2 \varv_\perp \right)
    \label{eq:v_coupled2}
    \ ,
\end{align}
which are coupled equations. In the special case where there is only variation
in the parallel direction Equation~\eqref{eq:v_coupled1} however reduces to
\be
    \pder{\varv_\para}{t} = \f{4\nu_\para}{3}\partial_\para^2 \varv_\para \ ,
    \label{eq:only_parallel}
\en
and, conversely, Equation~\eqref{eq:v_coupled2} reduces to
\be
    \pder{\varv_\perp}{t} = \f{\nu_\para}{3} \partial_\perp^2 \varv_\perp \ ,
\en
if there is only variation in the perpendicular direction. In these limiting
cases the problem thus reduces to solving a 1D diffusion equation. This is
useful for testing purposes as the 1D diffusion equation has a known
analytical solution.

The entropy equation gives the following evolution equation
\be
    \pder{\varepsilon}{t} =
    -\mathbf{\Pi} \vec{:} \del \vec{\varv} =
    \f{(\Delta p)^2}{3\rho \nu_\para}
    \label{eq:internal-energy-parallel}
\en
for the internal energy $\varepsilon = p/(\gamma-1)$, assuming that the
initial internal energy is uniform.

%-----------------------------------------------------------------------------
\subsection{1D solution used as test}
\label{sec:1d-diff-solution}

Consider an initial condition with velocity
\be
    \vec{\varv}(\vec{r}) = A \sin(\vec{k} \cdot \vec{r}) \b \ ,
\en
where $A$ is the amplitude of the perturbation and $\vec{r}=x\ex+y\ey+z\ez$ is
a position vector. If the
magnetic field, $\vec{B}$, and wave vector, $\vec{k}$, are aligned,
$\vec{k}=k_\para \b$, then the initial velocity is directed along the magnetic
field and only varies in magnitude along this direction. The
solution is therefore found by integrating Equation~\eqref{eq:only_parallel}.
We find
\begin{align}
    \varv_\para(\vec{x}, t) &= A \sin(\vec{k} \cdot \vec{r})\,\ue^{-\gamma t}
    \b \ ,
\end{align}
with the decay rate
\be
    \gamma = \f{4\nu_\para}{3}k^2 \ .
\en
The heating rate given by Equation~\eqref{eq:heating-term} leads to a time
evolution given by
\begin{align}
    \varepsilon(\vec{r}, t) &= \varepsilon_0 + \f{\rho A^2}{2} \cos^2(\vec{k}
    \cdot
    \vec{r}) \left(1 - \ue^{-2\gamma t}\right) \ ,
\end{align}
found by simple integration.
Here we have assumed that the internal energy is initially constant in space
with $\varepsilon(t=0)=\varepsilon_0$.

%-----------------------------------------------------------------------------
\subsection{General solution}

If there is variation in both parallel and perpendicular directions we can
proceed by Fourier transforming Equations~\eqref{eq:v_coupled1}
and \eqref{eq:v_coupled2} by assuming that the domain is periodic.
We obtain
\begin{align}
    \pder{\tilde{\varv}_\para}{t} &= -\f{2\nu_\para}{3}
    \left(2k_\para^2\tilde{\varv}_\para -
    k_\para k_\perp \tilde{\varv}_\perp \right) \ , \\
    \pder{\tilde{\varv}_\perp}{t} &= \f{\nu_\para}{3} \left(2k_\perp k_\para
    \tilde{\varv}_\para - k_\perp^2 \tilde{\varv}_\perp \right) \ ,
\end{align}
which is still a set of coupled equations. Consider however, instead the
evolution\footnote{This transformation assumes that both $k_\para$ and
$k_\perp$ are not zero. If one of them is zero, a solution can be found
directly without the need for a transformation, i.e., as outlined above.} of
$\xi = k_\perp/2k_\para \tilde{\varv}_\para + \tilde{\varv}_\perp$ and $\zeta
= - 2k_\para/k_\perp \tilde{\varv}_\para + \tilde{\varv}_\perp$. We then find
two decoupled equations
\begin{align}
    \pder{\xi}{t} &= 0 \ ,
    \label{eq:v1-equation}\\
    \pder{\zeta}{t} &= -\gamma \zeta \ ,
    \label{eq:v2-equation}
\end{align}
where
\be
    \gamma = \f{\nu_\para}{3}\left(4k_\para^2 + k_\perp^2\right) \ .
\en
The solutions to Equations~\eqref{eq:v1-equation} and \eqref{eq:v2-equation}
are thus
simply $\xi(t)
= \xi_0$ and
\be
    \zeta(t) = \zeta_0 \ue^{-\gamma t}\ ,
\en
where $\xi_0 = \xi(0)$ and $\zeta_0 = \zeta(0)$. The solutions for
$\tilde{\varv}_\para$ and $\tilde{\varv}_\perp$ are then found to be
\begin{align}
    \label{eq:decay2-vpar}
    \tilde{\varv}_\para(t) &= \f{2k_\para k_\perp
    \left(\xi_0 - \zeta_0 \ue^{-\gamma t} \right)}{4k_\para^2 + k_\perp^2} \ ,
    \\
    \label{eq:decay2-vperp}
    \tilde{\varv}_\perp(t) &=  \f{4 k_\para^2 \xi_0 + k_\perp^2\zeta_0
        \ue^{-\gamma t}}{4k_\para^2 + k_\perp^2} \ .
\end{align}
The Cartesian solution is then $\vec{\tilde{\varv}} = \tilde{\varv}_\para \b +
\tilde{\varv}_\perp \vec{n}$ where $\vec{n}$ is a unit vector normal to $\b$.

%-----------------------------------------------------------------------------
\subsection{2D solution used as test}
\label{sec:2d-diff-solution}

The test presented in Section~\ref{sec:decay2} has $\b = (\ex + \ey)/\sqrt{2}$
with variation in $x$ only. We thus have the special case $k_\para=k_\perp =
k/\sqrt{2}$ of the previous section for each mode $n$.
The solution given by Equations~\eqref{eq:decay2-vpar} and
\eqref{eq:decay2-vperp} was for a single mode. For a sum of modes we have
\begin{align}
    \varv_\para(t) &= \sum_n \f{2 \left[\xi_n(0) -
                \zeta_n(0) \ue^{-\gamma_n t} \right]}{5} \ , \\
    \varv_\perp(t) &= \sum_n  \f{4 \xi_n(0) +
     \zeta_n(0) \ue^{-\gamma_n t}}{5} \ ,
\end{align}
where
\be
\gamma_n = \f{5\nu_\para}{3} k_n^2 \ .
\en
The variables $\xi_n$ and $\zeta_n$ are related to $\varv_{\para, n}$ and
$\varv_{\perp, n}$ by $\xi_n = \varv_{\para, n}/2 + \varv_{\perp, n}$
and $\zeta_n = -2\varv_{\para, n} + \varv_{\perp, n}$.
Since $\varv_\para = \varv_y/\sqrt{2}$
and $\varv_\perp = -\varv_y/\sqrt{2}$ at $t=0$ it follows that
\begin{align}
    \xi_n(0) &= -c \f{a_n}{2\sqrt{2}}\cos(k_x x) \ , \\
    \zeta_n(0) &= -c \f{6a_n}{2\sqrt{2}} \cos(k_x x) \ ,
\end{align}
where $a_n$ are the Fourier coefficients for the initial $y$-velocity given in
Equation~\eqref{eq:fourier-an}. We thus find the solution for parallel and
perpendicular velocity profiles
\begin{align}
    \varv_\para(t) &= -c\sum_n \f{1 - 6 \ue^{-\gamma_n t}}{5\sqrt{2}}
         a_n \cos(k_n x) \ , \\
    \varv_\perp(t) &= -c\sum_n  \f{2 + 3 \ue^{-\gamma_n t}}{5\sqrt{2}}
         a_n \cos(k_n x)\ ,
\end{align}
which enables us to obtain $\varv_x(t) = (\varv_\para(t) +
\varv_\perp(t))/\sqrt{2}$ and $\varv_y(t) = (\varv_\para(t) -
\varv_\perp(t))/\sqrt{2}$ as given in Equations~\eqref{eq:vx-h-sol} and
\eqref{eq:vy-h-sol}, respectively.

The pressure anisotropy for this solution can be found using
Equation~\eqref{eq:pa-in-parallel} and the expressions for $\varv_\para$ and
$\varv_\perp$. We find that the pressure anisotropy is given by
Equation~\eqref{eq:pa-h-sol}, i.e.,
\be
    \Delta p(x, t) = -\f{3\rho \cs \nu_\para}{2} \sum_{n=1}^{\infty} k_n a_n
    \sin(k_n x)
    \,\ue^{-\gamma_n t} \ . \\
\en
The associated heating rate is evaluated using
Equation~\eqref{eq:internal-energy-parallel} which yields
\begin{align}
    &\pder{\varepsilon}{t} = \f{3\rho c^2 \nu_\para}{4}
                \left(\sum_n a_n k_n \sin(k_n x) \ue^{-\gamma_n t}\right)^2
                = \nonumber \\
       &         \f{9\rho c^2}{10}
        \sum_n\sum_m a_n a_m \sqrt{\gamma_n \gamma_m} \sin(k_n x) \sin(k_m x)
        \,\ue^{-(\gamma_n+\gamma_m) t}
        \label{eq:depsdt-parallel-sol}
\end{align}
where the identity $k_n k_m = \sqrt{\gamma_n \gamma_m} 6/(5\nu_\para)$
was used to simplify the double sum in the last step.
Integration of Equation~\eqref{eq:depsdt-parallel-sol} with respect to time
then gives $\varepsilon(x, t)$ as written in Equation~\eqref{eq:eps-h-sol}.

%-----------------------------------------------------------------------------
%-----------------------------------------------------------------------------
\section{Interruption by the firehose instability}
\label{sec:fft-calc}

We derive a set of two coupled wave equations which model the Alfv\'{e}n wave
interruption test that we present in Section~\ref{sec:firehose}. We use the
same geometry as in \citet{Squire2017}, i.e., we take the mean field to be in
the $z$-direction, the perpendicular perturbation to be in the $x$-direction
and variation to be in the $z$-direction only.

We have $\vec{B} = B_0 \ez +
\delta \vec{B}$ with $\delta \vec{b} = \delta \vec{B}/B_0 = \delta b \ex$
and the velocity
$\vec{\varv} = \varv \ex$. The induction equation, Equation~\eqref{eq:ind},
then yields
\be
    \pder{\delta b}{t} = \pder{\varv}{z} \ .
\en
The momentum equation becomes
\be
    \pder{\varv}{t} = \va^2 \pder{\delta b}{z} +
    \f{1}{\rho}\pder{}{z}\left(\f{\Delta p}{1 + \delta b^2}
    \delta b \right) \ ,
    \label{eq:squire-mom-tmp}
\en
where, since $\rho$ is constant, Equation~\eqref{eq:p-aniso} yields
\be
    \Delta p(z, t) = 3\rho \nu_\para \der{\ln B}{t} \ .
\en
As
\be
    B(t) = B_0 \sqrt{1 + \delta b^2} \ ,
\en
we find that
\be
    \Delta p(z, t) = \f{3\rho \nu_\para}{1 + \delta b^2}
    \delta b \pder{\delta b}{t}
    =
    \f{3\rho \nu_\para}{1 + \delta b^2}
    \pder{\varv}{z} \delta b  \ ,
    \label{eq:squire-pa}
\en
where the induction equation was used in the last step. Combining the
expression for $\Delta p$ given by Equation~\eqref{eq:squire-pa} and
Equation~\eqref{eq:squire-mom-tmp} gives us two coupled equations
\begin{align}
    \pder{\delta b}{t} &= \pder{\varv}{z} \ , \\
    \pder{\varv}{t} &= \va^2 \pder{\delta b}{z} +
    \pder{}{z}\left(\f{3 \nu_\para}{\left(1 + \delta b^2\right)^2}
    \pder{\varv}{z}\delta b^2 \right) \ ,
\end{align}
for $\varv$ and $\delta b$. We solve these coupled equations with the same
method as \citet{Squire2016,Squire2017}, i.e., spectral derivatives with the
Fast-Fourier transform (FFT), hyper-viscous damping of grid scale noise with a
$k^6$ diffusion operator and a fourth order Runge-Kutta update in time.

Our coupled equations differ, albeit only slightly, from the ones in
\citet{Squire2017} because we include the $1/(1+\delta b^2)$ nonlinearity in
Equation~\eqref{eq:squire-pa} for $\Delta p$. \citet{Squire2017}, who
considered small perturbations deliberately only kept it in the momentum
equation (Squire, private communication). The inclusion of the nonlinearity in
the expression for $\Delta p$ turns out to be necessary for a match between
the solutions obtained here and with \textsc{Athena} and \textsc{Arepo} (shown
in Fig.~\ref{fig:lin-polarized}).

% Don't change these lines
\bsp	% typesetting comment
\label{lastpage}
\end{document}